\documentclass[twoside,twocolumn,9pt]{article}
\usepackage{extsizes}
\usepackage[super,sort&compress,comma]{natbib} 
\usepackage[version=3]{mhchem}
\usepackage[left=1.5cm, right=1.5cm, top=1.785cm, bottom=2.0cm]{geometry}
\usepackage{balance}
\usepackage{widetext}
\usepackage{times,mathptmx}
\usepackage{sectsty}
\usepackage{graphicx} 
\usepackage{lastpage}
\usepackage[format=plain,justification=raggedright,singlelinecheck=false,font={stretch=1.125,small,sf},labelfont=bf,labelsep=space]{caption}
\usepackage{float}
\usepackage{fancyhdr}
\usepackage{fnpos}
\usepackage[english]{babel}
\usepackage{array}
\usepackage{droidsans}
\usepackage{charter}
\usepackage[T1]{fontenc}
\usepackage[usenames,dvipsnames]{xcolor}
\usepackage{setspace}
\usepackage[compact]{titlesec}

\definecolor{cream}{RGB}{222,217,201}

\DeclareMathOperator{\sign}{sign}

\begin{document}

\pagestyle{fancy}
\thispagestyle{plain}
\fancypagestyle{plain}{

\fancyhead[C]{\includegraphics[width=18.5cm]{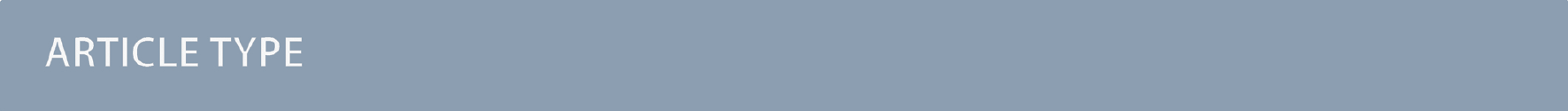}}
\fancyhead[L]{\hspace{0cm}\vspace{1.5cm}\includegraphics[height=30pt]{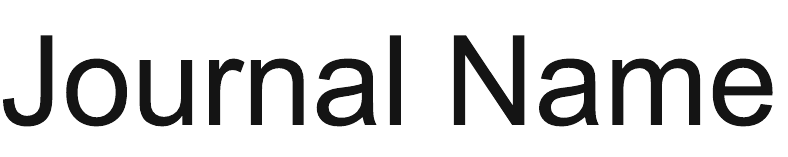}}
\fancyhead[R]{\hspace{0cm}\vspace{1.7cm}\includegraphics[height=55pt]{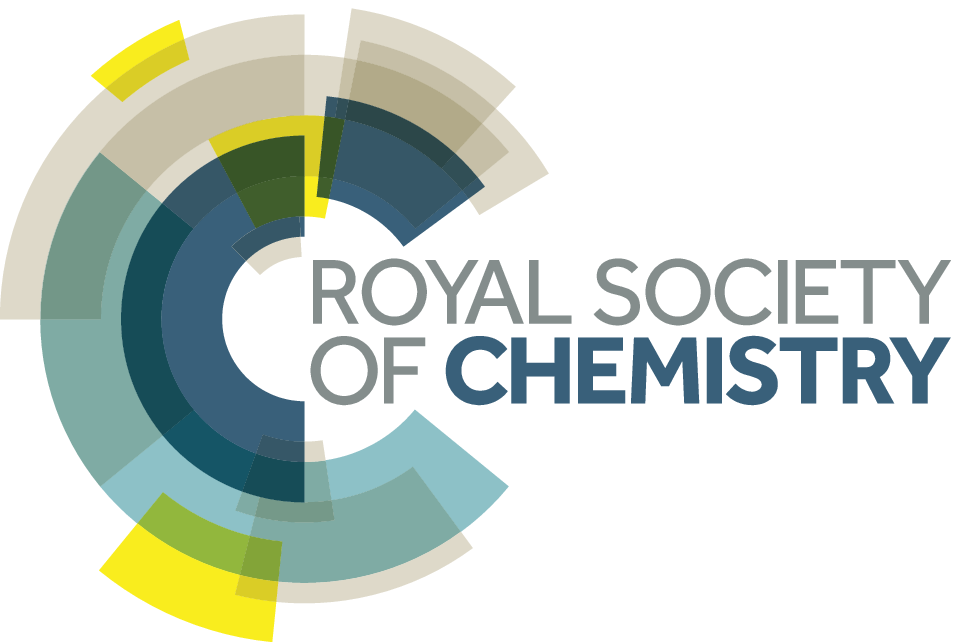}}
\renewcommand{\headrulewidth}{0pt}
}

\makeFNbottom
\makeatletter
\renewcommand\LARGE{\@setfontsize\LARGE{15pt}{17}}
\renewcommand\Large{\@setfontsize\Large{12pt}{14}}
\renewcommand\large{\@setfontsize\large{10pt}{12}}
\renewcommand\footnotesize{\@setfontsize\footnotesize{7pt}{10}}
\makeatother

\renewcommand{\thefootnote}{\fnsymbol{footnote}}
\renewcommand\footnoterule{\vspace*{1pt}%
\color{cream}\hrule width 3.5in height 0.4pt \color{black}\vspace*{5pt}} 
\setcounter{secnumdepth}{5}

\makeatletter 
\renewcommand\@biblabel[1]{#1}            
\renewcommand\@makefntext[1]%
{\noindent\makebox[0pt][r]{\@thefnmark\,}#1}
\makeatother 
\renewcommand{\figurename}{\small{Fig.}~}
\sectionfont{\sffamily\Large}
\subsectionfont{\normalsize}
\subsubsectionfont{\bf}
\setstretch{1.125} 
\setlength{\skip\footins}{0.8cm}
\setlength{\footnotesep}{0.25cm}
\setlength{\jot}{10pt}
\titlespacing*{\section}{0pt}{4pt}{4pt}
\titlespacing*{\subsection}{0pt}{15pt}{1pt}

\fancyfoot{}
\fancyfoot[LO,RE]{\vspace{-7.1pt}\includegraphics[height=9pt]{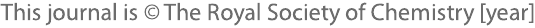}}
\fancyfoot[CO]{\vspace{-7.1pt}\hspace{13.2cm}\includegraphics{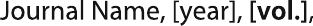}}
\fancyfoot[CE]{\vspace{-7.2pt}\hspace{-14.2cm}\includegraphics{head_foot/RF}}
\fancyfoot[RO]{\footnotesize{\sffamily{1--\pageref{LastPage} ~\textbar  \hspace{2pt}\thepage}}}
\fancyfoot[LE]{\footnotesize{\sffamily{\thepage~\textbar\hspace{3.45cm} 1--\pageref{LastPage}}}}
\fancyhead{}
\renewcommand{\headrulewidth}{0pt} 
\renewcommand{\footrulewidth}{0pt}
\setlength{\arrayrulewidth}{1pt}
\setlength{\columnsep}{6.5mm}
\setlength\bibsep{1pt}

\makeatletter 
\newlength{\figrulesep} 
\setlength{\figrulesep}{0.5\textfloatsep} 

\newcommand{\topfigrule}{\vspace*{-1pt}%
\noindent{\color{cream}\rule[-\figrulesep]{\columnwidth}{1.5pt}} }

\newcommand{\botfigrule}{\vspace*{-2pt}%
\noindent{\color{cream}\rule[\figrulesep]{\columnwidth}{1.5pt}} }

\newcommand{\dblfigrule}{\vspace*{-1pt}%
\noindent{\color{cream}\rule[-\figrulesep]{\textwidth}{1.5pt}} }

\makeatother

\twocolumn[
  \begin{@twocolumnfalse}
\vspace{3cm}
\sffamily
\begin{tabular}{m{4.5cm} p{13.5cm} }

\includegraphics{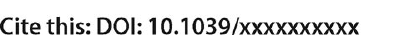} & \noindent\LARGE{\textbf{Theory of defect motion in 2D passive and active nematic liquid crystals}} \\
\vspace{0.3cm} & \vspace{0.3cm} \\

 & \noindent\large{Xingzhou Tang\textit{$^{a}$} and Jonathan V. Selinger$^{\ast}$\textit{$^{a}$}} \\

\includegraphics{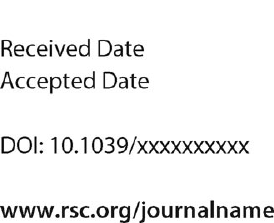} & \noindent\normalsize{The motion of topological defects is an important feature of the dynamics of all liquid crystals, and is especially conspicuous in active liquid crystals.  Understanding defect motion is a challenging theoretical problem, because the dynamics of orientational order is coupled with backflow of the fluid, and because a liquid crystal has several distinct viscosity coefficients.  Here, we suggest a coarse-grained, variational approach, which describes the motion of defects as effective ``particles.''  For passive liquid crystals, the theory shows how the drag depends on defect orientation, and shows the coupling between translational and rotational motion.  For active liquid crystals, the theory provides an alternative way to describe motion induced by the activity coefficient.} \\

\end{tabular}

 \end{@twocolumnfalse} \vspace{0.6cm}

  ]

\renewcommand*\rmdefault{bch}\normalfont\upshape
\rmfamily
\section*{}
\vspace{-1cm}


\footnotetext{\textit{$^{a}$~Department of Physics and Liquid Crystal Institute, Kent State University, Kent, OH 44242, USA; E-mail: jselinge@kent.edu}}





\section{Introduction}

One important feature of the dynamics of liquid crystals is the motion of topological defects.  In conventional, passive liquid crystals, defects form when a disordered phase is quenched into a more ordered phase, e.g., when isotropic is quenched into nematic, or smectic-A into smectic-C.  After the quench, defects of opposite topological charge move together and annihilate each other.\cite{Chuang1991,Bowick1994,Pargellis1991,Pargellis1992,Oswald2005,Blanc2005,Stannarius2006,Dierking2012,Guimaraes2013,Kim2013,Stannarius2016}  Their motion is driven by the interaction among defects, as well as by boundary conditions and applied fields.  In active liquid crystals, defects are constantly in motion, forming and annihilating each other, driven by the activity of the underlying medium.\cite{Sanchez2012,Keber2014,DeCamp2015}  In particular, defects of topological change $+1/2$ move with a characteristic velocity, while defects of topological charge $-1/2$ move diffusively.

To model the motion of topological defects, researchers have used two types of theoretical approaches.  First, and most fundamentally, one can use hydrodynamic equations to describe the simultaneous evolution of the liquid crystal order and the flow velocity fields throughout the system.  For passive liquid crystals, the hydrodynamic equations can be derived from Ericksen-Leslie theory expressed in terms of the director field,\cite{Ericksen1960,Ericksen1961,Leslie1966,Leslie1968} or from Beris-Edwards theory expressed in terms of the nematic order tensor.\cite{Beris1994}  These equations can be solved numerically to obtain the liquid crystal order and flow velocity as functions of position and time, and these solutions can include the motion of defects.\cite{Toth2002,Svensek2003}  For active liquid crystals, one can likewise construct hydrodynamic equations, which include an extra term representing the activity.\cite{Simha2002,Marchetti2013,Prost2015,Ramaswamy2017,Doostmohammadi2018}  When these equations are solved for the liquid crystal order and flow velocity, they show the formation, motion, and annihilation of defects.

As an alternative theoretical approach, one can model defects as if they were effective ``particles,''  which move in response to the total forces acting on them.  This approach is more coarse-grained than hydrodynamics, because it describes the motion in terms of just a few degrees of freedom for each defect, while hydrodynamics describes liquid crystal order and flow velocity at every point in the system.  The forces on defects have been investigated by several researchers over many years, in the context of passive liquid crystals.  The elastic force was derived in a classic calculation,\cite{Dafermos1970} which shows that the interaction energy scales logarithmically with the separation $r$ between defects, and hence the force scales as $1/r$, in two dimensions (2D).  The drag force was first derived through a simple theory, which assumes small defect velocity and neglects fluid flow, and thereby predicts a drag coefficient diverging logarithmically with system size.\cite{Imura1973}  Further studies considered the possibility of larger defect velocity, so that the divergence with system size is cut off by a velocity-dependent length scale, leading to anomalous scaling of the drag with velocity.\cite{Pismen1990,Ryskin1991,Denniston1996,Radzihovsky2015}  Other models included fluid flow, and hence found more complex results for the drag, which is different for positive and negative topological defects.\cite{Kats2002,Sonnet2005,Sonnet2009}

In the context of active liquid crystals, several papers have used the effective particle approach to predict the statistical mechanics of defect formation, motion, and annihilation.\cite{Giomi2013,Pismen2013,Giomi2014,Zhang2018,Cortese2018}  This approach has been generalized to the motion of topologically required defects on the surface of a sphere.\cite{Keber2014}  Those studies have shown that active defects should not be regarded as just point particles, but rather as \emph{oriented particles}.  In particular, defects of topological charge $+1/2$ are surrounded by a comet-shaped director field, and the orientation of the comet determines the direction of self-propelled motion, while defects of topological charge $-1/2$ are surrounded by a triangular director field.  Moreover, experimental and numerical studies of systems with many defects have found statistical order in the defect orientations.\cite{DeCamp2015}  Motivated by those results, Vromans and Giomi developed a formalism to describe defect orientations by vectors,\cite{Vromans2016} and we generalized the formalism using tensors.\cite{Tang2017}  Most recently, Shankar \emph{et al}.\ derived the general orientational dynamics of defects in active liquid crystals, and used those results to predict the nonequilibrium defect unbinding transition.\cite{Shankar2018}

The first purpose of this paper, in Section~2, is to apply the concept of defect orientation to the dynamics of \emph{passive} liquid crystals. In our previous paper about defect orientations, we determined the effect of orientation on the elastic interaction between defects in passive liquid crystals.  This interaction generates elastic forces and torques on the defects.  Now, we investigate the effect of orientation on drag forces and torques.  We determine what translational and orientational drag coefficients are allowed by the symmetry of defects, and assess how translational drag coefficients depend the relative angle between defect orientation and velocity.

In the course of doing this calculation, we develop a formalism for defect motion in passive liquid crystals based on the Rayleigh dissipation function.  We suggest that this formalism is particularly useful for coarse-graining the dissipative dynamics, from the hydrodynamic level to the effective particle level, because the same dissipation function can be expressed on either length scale.  On the hydrodynamic level, it can be written in terms of the liquid crystal order and the flow velocity fields, with Ericksen-Leslie viscosity coefficients.  Similarly, on the effective particle level, it can be written in terms of symmetry-allowed combinations of the defect velocity and orientation vectors, with effective drag coefficients.  By comparing these expressions, we can determine how the effective drag coefficients are related to Ericksen-Leslie viscosity coefficients.

One result of this calculation is that some Ericksen-Leslie viscosities give drag that is independent of defect orientation, while other viscosities give drag that depends on defect orientation.  Another result is that the drag on a positive topological charge is less than drag on a negative topological charge because of backflow effects, and hence positive topological charges move more rapidly, in agreement with experiments\cite{Oswald2005,Blanc2005,Dierking2012} and previous calculations using other methods.\cite{Toth2002,Svensek2003,Kats2002,Sonnet2005,Sonnet2009}  We provide an example of how this method can be used to predict the motion of a defect in a channel, driven by boundary conditions.

In Section~3, we apply the formalism based on the Rayleigh dissipation function back to active liquid crystals.  This calculation shows that activity can be represented by one extra term in the dissipation function, either on the hydrodynamic level or the effective particle level.  Although this term is not positive-definite, and hence is not exactly a dissipation, it plays the role of the Rayleigh dissipation function in the equations of motion.  Because of this term, $+1/2$ defects move with a velocity proportional to the activity coefficient, in the direction given by the defect orientation vector.  We construct two examples of how this method can predict the motion of a defect, driven by activity.  We recognize that these results for active liquid crystals are not new; they have been found through other approaches by Shankar \emph{et al}.\cite{Shankar2018} and previous articles.  Even so, we think it is useful to present them here, using the formalism of the Rayleigh dissipation function and defect orientation vector, because we find this approach to be intuitive and other investigators might also.

Finally, in Section~4, we discuss these results, and consider the prospects for extending them to other defects and textures in passive and active liquid crystals.

\section{Passive liquid crystals}

\subsection{Statement of problem}

In this work, we consider a 2D nematic liquid crystal.  At each point in the material, there is some orientational order, which may be described by the director field $\hat{\mathbf{n}}(\mathbf{r},t)$ or the nematic order tensor $Q_{ij}(\mathbf{r},t)$, as well as a fluid flow velocity $\mathbf{v}(\mathbf{r},t)$.  A full description of the dynamics must involve coupled partial differential equations for orientational order and fluid flow velocity.  Solving these equations is a complex problem, which usually can only be done numerically.  Our goal is to provide a coarse-grained description of the dynamics in terms of a reduced number of degrees of freedom associated with topological defects.

\begin{figure}
\includegraphics[width=.49\columnwidth]{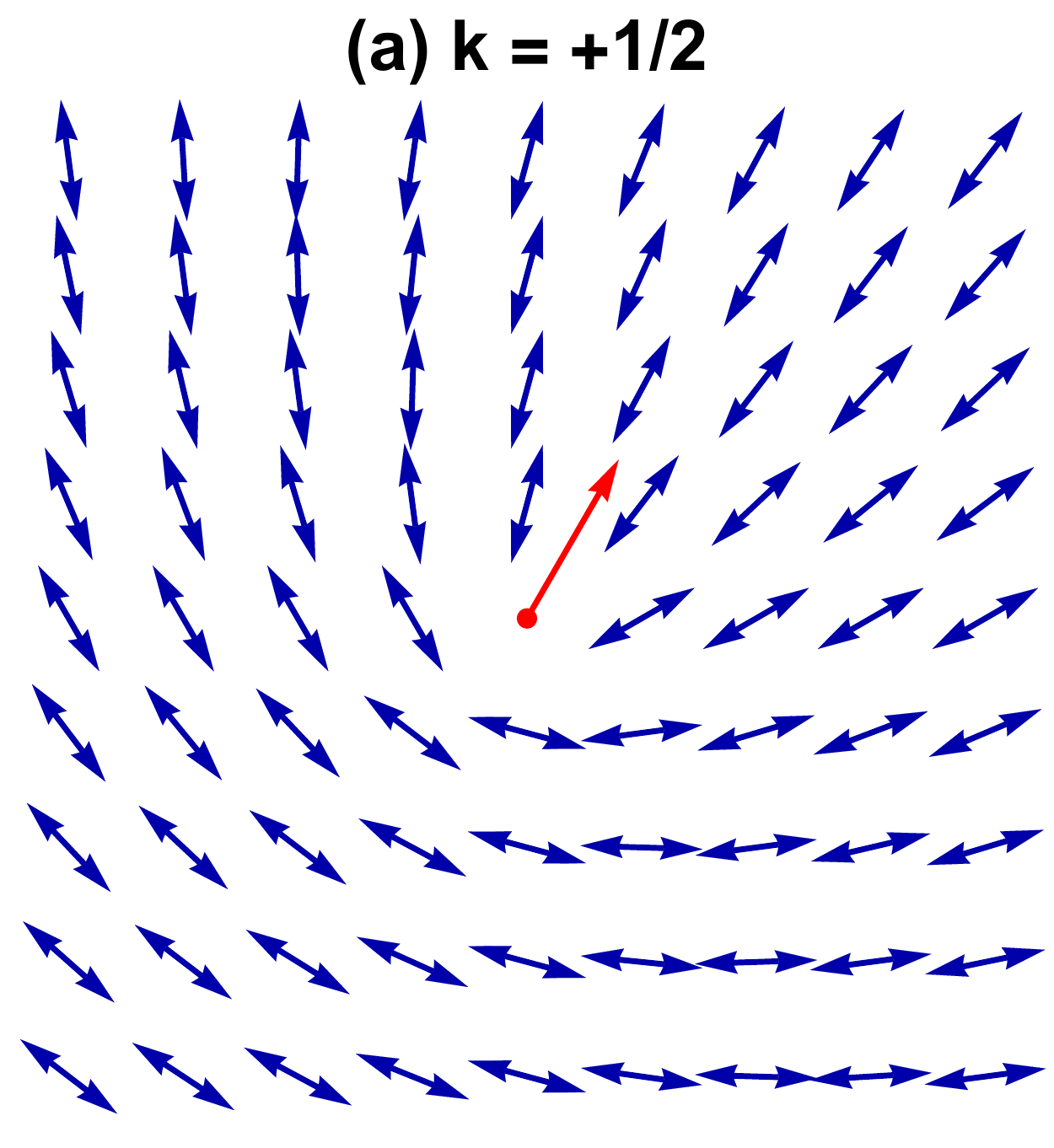}\hfill\includegraphics[width=.49\columnwidth]{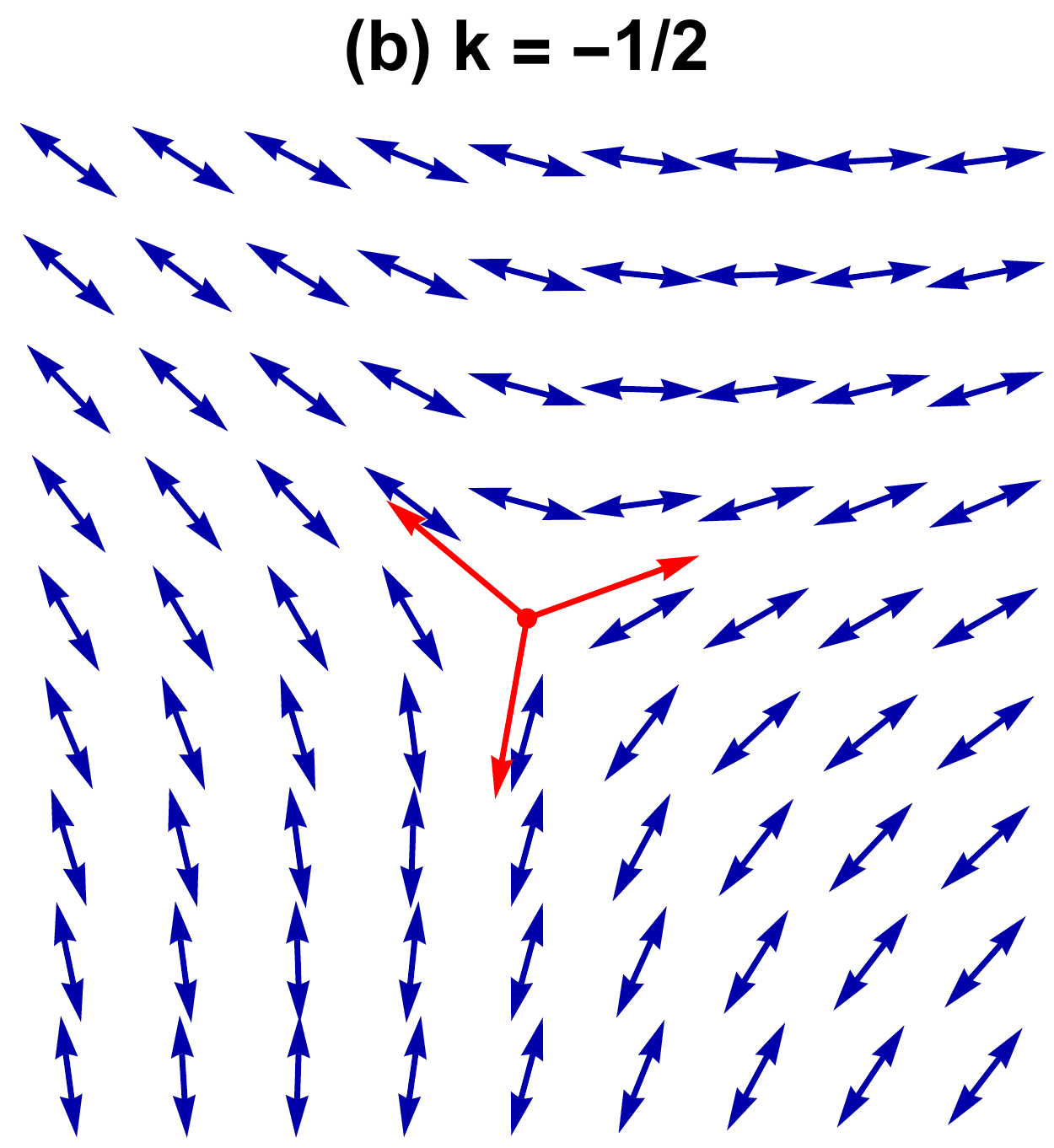}
\caption{Examples of defects in a 2D nematic liquid crystal, with red arrows indicating the defect orientation.}
\end{figure}

Suppose the liquid crystal has a topological defect at position $\mathbf{R}(t)=(X(t),Y(t))$.  This defect is characterized by a topological charge $k$, which is a half-integer or integer indicating how many times the director rotates as one passes through a loop around the defect.  As discussed in recent papers,\cite{Vromans2016,Tang2017} the defect is \emph{also} characterized by an orientation, which describes where the director points radially outward from (or inward toward) the defect.  This orientation is defined up to rotations through an angle of $\pi/|1-k|$.  Hence, as we argued previously,\cite{Tang2017} the defect orientation in a nematic phase can be represented by a tensor of rank $2|1-k|$.  For topological charge $k=+1/2$, the defect orientation is a unit vector $\mathbf{p}(t)=(\cos\Psi(t),\sin\Psi(t))$, as illustrated by the arrow in Fig.~1(a).  For topological charge $k=-1/2$, the defect orientation is a third-rank, completely symmetric tensor $T_{ijk}(t)$, with $T_{xxx}=-T_{xyy}=-T_{yxy}=-T_{yyx}=\frac{1}{2}\cos3\Psi$ and $T_{xxy}=T_{xyx}=T_{yxx}=-T_{yyy}=\frac{1}{2}\sin3\Psi$, as represented by the triad of arrows in Fig.~1(b).  Hence, the coarse-grained description should provide equations of motion for $X(t)$, $Y(t)$, and $\Psi(t)$.

For the \emph{static} physics, there is a well-established procedure to go from the microscopic theory based on the director field $\hat{\mathbf{n}}(\mathbf{r},t)$ to the coarse-grained theory based on the defect degrees of freedom.  In this procedure, one minimizes the Frank free energy, subject to the constraint that topological defects are at specified positions, and determines how the minimum free energy $F$ depends on the defect positions.  This dependence gives an effective interaction of defects with boundary conditions or with other defects.  Thus, a classic calculation shows that the interaction energy scales logarithmically with the separation between defects.\cite{Dafermos1970}  In our previous paper,\cite{Tang2017} we generalized this procedure to include defect orientation, and found that there is an extra interaction energy if the defects do not have the optimal relative orientation.  Hence, the elastic force acting on the position of a defect is $\mathbf{f}_\text{elastic}=-\partial F/\partial{\mathbf{R}}$, and the elastic force acting on the defect orientation is $f_\text{elastic}=-\partial F/\partial\Psi$.

For the \emph{dynamic} physics, we need a procedure to go from the microscopic theory based on the Ericksen-Leslie equations for $\hat{\mathbf{n}}(\mathbf{r},t)$ and $\mathbf{v}(\mathbf{r},t)$ to a coarse-grained theory based on defect degrees of freedom.  Here, we suggest an approach using the Rayleigh dissipation function, which is a theoretical construction representing half the rate of dissipating mechanical energy into heat.

In most theoretical work, the fundamental hydrodynamic theory is expressed in terms of the stress tensor.  However, an alternative formulation of the same theory is expressed in terms of the Rayleigh dissipation function.  To our knowledge, this version of the theory was first suggested by Vertogen,\cite{Vertogen1983,Vertogen1988} and related ideas have been advocated by Sonnet and Virga\cite{Sonnet2001,Sonnet2012} and by Doi.\cite{Doi2011}  This approach begins by listing all of the modes that dissipate energy.  Next, the Rayleigh dissipation function is constructed as the most general scalar that is allowed by symmetry, at quadratic order in these modes.  Finally, the drag forces are found by differentiating the dissipation function with respect to the generalized velocities.

On a microscopic basis, there are two modes that dissipate energy:  the strain rate tensor, $A_{ij}=\frac{1}{2}(\partial_i v_j + \partial_j v_i)$, and the director rotation with respect to the background fluid vorticity, $N_i = \dot{n}_i - \frac{1}{2}(\partial_j v_i - \partial_i v_j)n_j$.  In terms of these two modes, the most general quadratic dissipation function can be constructed as\cite{Stewart2004}
\begin{align}
D=\int d^2 r \biggl[&\frac{1}{2}\alpha_4 A_{ij}A_{ij} + \frac{1}{2}(\alpha_5 + \alpha_6)n_i A_{ij}A_{jk}n_k + \frac{1}{2}\alpha_1 (n_i A_{ij}n_j)^2\nonumber\\
&+\frac{1}{2}\gamma_1 N_i N_i + \gamma_2 N_i A_{ij} n_j\biggr].
\label{dissipationfull}
\end{align}
Here, the $\alpha$ coefficients are the Leslie viscosities for fluid flow.  Note that $\alpha_4$ is the isotropic viscosity, while the other terms provide corrections depending on the direction of the strain rate with respect to the director.  For a 2D incompressible flow (unlike 3D), we have the identity $2n_i A_{ij}A_{jk}n_k = A_{ij}A_{ij}$, and hence the second term is equivalent to the first.\cite{Ryskin1991b}  By comparison, $\gamma_1$ is the rotational viscosity for director rotation with respect the the background fluid vorticity.  Finally, $\gamma_2$ is the torsion coefficient, which expresses a dissipative coupling between strain rate and director rotation.

On a macroscopic basis, we can repeat the same type of analysis based purely on symmetry considerations.  For a $+1/2$ defect, there are two modes that dissipate energy:  the translational velocity $\mathbf{\dot{R}}$ and the rotational velocity $\mathbf{\dot{p}}$.  In terms of those two modes, the most general quadratic dissipation function can be constructed as
\begin{equation}
D=\frac{1}{2} D_1 |\mathbf{\dot{R}}|^2 + \frac{1}{2} D_2 (\mathbf{p}\cdot\mathbf{\dot{R}})^2 + \frac{1}{2} D_3 |\mathbf{\dot{p}}|^2
+ D_4 \mathbf{\dot{p}}\cdot\mathbf{\dot{R}}.
\label{macroscopicdplushalf}
\end{equation}
Here, $D_1$ shows the energy dissipated by defect translation, and $D_2$ shows how that energy dissipation depends on the defect orientation with respect to the velocity.  Similarly, $D_3$ shows the energy dissipated by defect rotation, and $D_4$ shows a dissipative coupling between defect translation and rotation.  This quadratic form is positive-definite if $D_4^2 < D_1 D_3$.  The drag force acting on the defect position is
\begin{equation}
\mathbf{f}_\text{drag}=-\frac{\partial D}{\partial\mathbf{\dot{R}}}
=-D_1\mathbf{\dot{R}}-D_2\mathbf{p}(\mathbf{p}\cdot\mathbf{\dot{R}})-D_4\mathbf{\dot{p}},
\end{equation}
and the drag force acting on the defect orientation is
\begin{equation}
f_\text{drag}=-\frac{\partial D}{\partial\dot{\Psi}}=-D_3 \dot{\Psi} -D_4 \mathbf{p}\times\mathbf{\dot{R}}.
\end{equation}
Those forces can be combined into a matrix equation as
\begin{equation}
\begin{bmatrix}
f^\text{drag}_x\\
f^\text{drag}_y\\
f^\text{drag}_\Psi
\end{bmatrix}
=-
\begin{bmatrix}
D_1 + D_2 \cos^2 \Psi & D_2 \cos\Psi\sin\Psi & -D_4 \sin\Psi\\
D_2 \cos\Psi\sin\Psi & D_1 + D_2 \sin^2 \Psi & D_4 \cos\Psi\\
-D_4 \sin\Psi & D_4 \cos\Psi & D_3
\end{bmatrix}
\begin{bmatrix}
\dot{x}\\
\dot{y}\\
\dot{\Psi}
\end{bmatrix}.
\end{equation}
If a translational or rotational force is applied to the defect, the steady-state response is given by $f^\text{app}_i + f^\text{drag}_i = 0$, and hence
\begin{equation}
\begin{bmatrix}
\dot{x}\\
\dot{y}\\
\dot{\Psi}
\end{bmatrix}
=
\begin{bmatrix}
D_1 + D_2 \cos^2 \Psi & D_2 \cos\Psi\sin\Psi & -D_4 \sin\Psi\\
D_2 \cos\Psi\sin\Psi & D_1 + D_2 \cos^2 \Psi & D_4 \cos\Psi\\
-D_4 \sin\Psi & D_4 \cos\Psi & D_3
\end{bmatrix}^{-1}
\begin{bmatrix}
f^\text{app}_x\\
f^\text{app}_y\\
f^\text{app}_\Psi
\end{bmatrix}.
\label{plushalfmobility}
\end{equation}
Hence, a $+1/2$ defect responds to an applied force with a mobility tensor given by the inverse matrix in Eq.~(\ref{plushalfmobility}).  This mobility tensor has the same structure as that of a boomerang-shaped colloidal particle.\cite{Chakrabarty2013}  In particular, we note that a translational force can induce rotational motion, and a rotational force can induce translational motion.

Similar considerations apply to a $-1/2$ defect.  There are two modes that dissipate energy:  The translational velocity $\mathbf{\dot{R}}$ and the time derivative of the orientation tensor $\dot{T}_{ijk}$.  In terms of these modes, most general quadratic dissipation function becomes
\begin{equation}
D=\frac{1}{2} D'_1 |\mathbf{\dot{R}}|^2 + \frac{1}{2} D'_3 \dot{T}_{ijk} \dot{T}_{ijk}
=\frac{1}{2} D'_1 |\mathbf{\dot{R}}|^2 + \frac{9}{2} D'_3 \dot{\Psi}^2,
\label{macroscopicdminushalf}
\end{equation}
where $D'_1$ shows the dissipation due to defect translation and $D'_3$ shows the dissipation due to defect rotation.  At quadratic order, symmetry does not allow any couplings between translation and orientation.  Hence, the matrix equation for drag forces is simply
\begin{equation}
\begin{bmatrix}
f^\text{drag}_x\\
f^\text{drag}_y\\
f^\text{drag}_\Psi
\end{bmatrix}
=-
\begin{bmatrix}
D'_1 & 0 & 0\\
0 & D'_1 & 0\\
0 & 0 & 9D'_3
\end{bmatrix}
\begin{bmatrix}
\dot{x}\\
\dot{y}\\
\dot{\Psi}
\end{bmatrix},
\end{equation}
and the steady-state response to an applied force is
\begin{equation}
\begin{bmatrix}
\dot{x}\\
\dot{y}\\
\dot{\Psi}
\end{bmatrix}
=
\begin{bmatrix}
D'_1 & 0 & 0\\
0 & D'_1 & 0\\
0 & 0 & 9D'_3
\end{bmatrix}^{-1}
\begin{bmatrix}
f^\text{app}_x\\
f^\text{app}_y\\
f^\text{app}_\Psi
\end{bmatrix}.
\label{minushalfmobility}
\end{equation}
As a result, a $-1/2$ defect has the mobility tensor given by the inverse matrix in Eq.~(\ref{minushalfmobility}).  Because that tensor is diagonal, a translational force induces only translational motion, and a rotational force induces only rotational motion, at lowest order in the forces.

The matrix equations (\ref{plushalfmobility}) and (\ref{minushalfmobility}) can be used directly, with the macroscopic $D$ and $D'$ coefficients considered as purely phenomenological parameters.  However, one might want to determine these macroscopic coefficients in terms of the more microscopic $\alpha$ and $\gamma$ coefficients.  That is the purpose of our coarse-graining calculation in the following sections.

\subsection{Minimal model}

As a first step, we consider a defect moving at a specified velocity with a fixed orientation.  We want to calculate its dissipation from microscopic theory, and compare the result with the calculation from macroscopic theory.  Although we use a minimal model, the calculation is still rather lengthy.  Readers who are mainly interested in the result rather than the method may wish to skip ahead to Eq.~(\ref{minimalDresult}).

Our minimal model of a 2D nematic liquid crystal is analogous to the model of a hexatic liquid crystal considered by Kats \emph{et al}.\cite{Kats2002}  We make the approximation of equal Frank constants, so that the Frank free energy becomes
\begin{equation}
F=\int d^2 r\left[\frac{1}{2}K(\partial_i n_j)(\partial_i n_j)\right].
\end{equation}
Similarly, we consider just two viscosity coefficients, the isotropic fluid flow viscosity $\alpha_4$ and the rotational viscosity $\gamma_1$, so that the dissipation function becomes
\begin{equation}
D=\int d^2 r\left[\frac{1}{2}\alpha_4 A_{ij}A_{ij} +\frac{1}{2}\gamma_1 N_i N_i\right].
\end{equation}
The minimal model requires both $\alpha_4 >0$ and $\gamma_1 >0$, so that the system will have drag against shear flow and drag against director rotation.  The other viscosity coefficients represent more subtle anisotropies in the viscous drag, and they will be added later as perturbations.  Note that the limit of $\alpha_4\to\infty$ corresponds to orientational order in a material that cannot flow.

We write the director field as $\hat{\mathbf{n}}=(\cos\theta,\sin\theta)$, so that the Frank free energy simplifies to
\begin{equation}
F=\int d^2 r\left[\frac{1}{2}K|\mathbf{\nabla}\theta|^2\right].
\end{equation}
Also, we assume that the material is incompressible, which implies that $\partial_i v_i = 0$.  Because of this constraint, the velocity field can be written in terms of a stream function $\psi(\mathbf{r},t)$ as $v_i=\epsilon_{ij}\partial_j \psi$, where $\epsilon_{ij}$ is the 2D Levi-Civita symbol.  The stream function $\psi$ is a standard concept in fluid mechanics, and should not be confused with the defect orientation angle $\Psi$.  In terms of the stream function, the strain rate tensor becomes
\begin{equation}
A_{ij}=\frac{1}{2}(\partial_i v_j + \partial_j v_i)=\frac{1}{2}(\epsilon_{jk}\partial_i \partial_k \psi + \epsilon_{ik}\partial_j \partial_k \psi). 
\end{equation}
Likewise, the background fluid vorticity becomes $\omega=\frac{1}{2}\epsilon_{ij}\partial_i v_j =-\frac{1}{2}\nabla^2 \psi$, and the director rotation with respect to the background fluid becomes
\begin{align}
N_i&=\dot{n}_i-\omega\epsilon_{ji}n_j
=\partial_t n_i + v_k \partial_k n_i-\omega\epsilon_{ji}n_j \nonumber\\
&=\epsilon_{ji}n_j\left[\partial_t \theta +\epsilon_{kl}(\partial_k \theta)(\partial_l \psi)+\frac{1}{2}\nabla^2 \psi\right].
\end{align}
Here, the first term $\dot{n}_i$ becomes a convective derivative, which leads to the nonlinear coupling $(\partial_k \theta)(\partial_l \psi)$.  The dissipation function then simplifies to
\begin{align}
D=\int d^2 r\Biggl[&\frac{1}{2}\alpha_4 \left[(\partial_i \partial_j \psi)(\partial_i \partial_j \psi)-\frac{1}{2}(\nabla^2 \psi)^2 \right]\nonumber\\
&+\frac{1}{2}\gamma_1 \left[\partial_t \theta +\epsilon_{kl}(\partial_k \theta)(\partial_l \psi)+\frac{1}{2}\nabla^2 \psi\right]^2\Biggr].
\label{minimalDintermsofthetaandpsi}
\end{align}

From the free energy and the dissipation function, we can derive the equations of motion for $\theta$ and $\psi$.  For the director orientation $\theta$, the elastic force is $-\delta F/\delta\theta(\mathbf{r},t)$, and the drag force is $-\delta D/\delta[\partial_t \theta(\mathbf{r},t)]$.  Hence, the equation for overdamped motion is that the forces must sum to zero,
\begin{align}
0&=-\frac{\delta F}{\delta\theta(\mathbf{r},t)}-\frac{\delta D}{\delta[\partial_t \theta(\mathbf{r},t)]} \nonumber\\
&=K\nabla^2 \theta-\gamma_1 \left[\partial_t \theta +\epsilon_{kl}(\partial_k \theta)(\partial_l \psi)+\frac{1}{2}\nabla^2 \psi\right].
\label{equationtheta}
\end{align}
For the generalized velocity $\psi$, the elastic force is zero, and the drag force is $-\delta D/\delta\psi(\mathbf{r},t)$.  Hence, the equation for overdamped motion is that the drag force equals zero,
\begin{align}
\label{equationpsi}
0=&-\frac{\delta D}{\delta\psi(\mathbf{r},t)} 
=-\frac{1}{2}\alpha_4 \nabla^4 \psi \\
&\qquad\quad+\gamma_1 \left[\epsilon_{ij}(\partial_i \theta)\partial_j -\frac{1}{2}\nabla^2\right]\left[\partial_t \theta +\epsilon_{kl}(\partial_k \theta)(\partial_l \psi)+\frac{1}{2}\nabla^2 \psi\right]\nonumber.
\end{align}
These equations are nonlinear because of the convective derivative.  As a check, in the limit of high viscosity $\alpha_4\to\infty$, Eq.~(\ref{equationpsi}) implies that $\psi$ is constant, meaning that the material does not flow.  Equation~(\ref{equationtheta}) then becomes the standard diffusion equation $\gamma_1 \partial_t \theta = K \nabla^2 \theta$.

We seek a solution of these equations corresponding to steady motion of a defect with a specified velocity $\mathbf{u}$.  In this steady state, we have $\theta(\mathbf{r},t)=\theta(\mathbf{r}-\mathbf{u}t)$ and $\psi(\mathbf{r},t)=\psi(\mathbf{r}-\mathbf{u}t)$.  Hence, the time derivative becomes $\partial_t \theta=-u_k \partial_k \theta$, and the equations of motion take the time-independent form
\begin{align}
0=&K\nabla^2 \theta +\gamma_1 \left[u_k \partial_k \theta -\epsilon_{kl}(\partial_k \theta)(\partial_l \psi)-\frac{1}{2}\nabla^2 \psi\right],\\
0=&-\frac{1}{2}\alpha_4 \nabla^4 \psi \\
&-\gamma_1\left[\epsilon_{ij}(\partial_i \theta)\partial_j -\frac{1}{2}\nabla^2\right]\left[u_k \partial_k \theta -\epsilon_{kl}(\partial_k \theta)(\partial_l \psi)-\frac{1}{2}\nabla^2 \psi\right] .\nonumber
\end{align}
To solve these equations, we choose a coordinate system such that the defect velocity $\mathbf{u}$ is in the $x$-direction, with $\mathbf{u}=u\hat{\mathbf{x}}$.  We then assume that $u$ is small, so that we can use perturbation theory as in Pismen and Rodriguez,\cite{Pismen1990} writing
\begin{align}
\theta(\mathbf{r})&=\theta_0(\mathbf{r})+u\theta_1(\mathbf{r})+O(u^2),\nonumber\\
\psi(\mathbf{r})&=\psi_0(\mathbf{r})+u\psi_1(\mathbf{r})+O(u^2).
\end{align}
At zeroth order in $u$, we assume that $\psi_0(\mathbf{r})$ is constant, meaning that the material does not flow if the defect does not move.  With this assumption, the second differential equation is identically satisfied, and the first differential equation becomes Laplace's equation $0=K\nabla^2 \theta_0$.  The solution of this equation, corresponding to a defect at the origin, can be written as
\begin{equation}
\theta_0=k\tan^{-1}\left(\frac{y}{x}\right)+\Theta_0.
\label{unperturbedtheta}
\end{equation}
Here, $k$ is the topological charge of the defect, and $\Theta_0$ represents an overall rotation of the director about the $z$-axis.  Previous papers\cite{Vromans2016,Tang2017} have shown that $\Theta_0$ is related to the defect orientation $\Psi$  by $\Psi=\Theta_0/(1-k)$ (mod $\pi/|1-k|$).  In particular, for a defect of charge $k=+1/2$, we have $\Psi=2\Theta_0$.  For a defect of charge $k=-1/2$, we have $\Psi=\frac{2}{3}\Theta_0$.

At first order in $u$, the differential equations become
\begin{align}
0=&K\nabla^2 \theta_1+\gamma_1 \left[\partial_x \theta_0 -\epsilon_{kl}(\partial_k \theta_0)(\partial_l \psi_1)-\frac{1}{2}\nabla^2 \psi_1\right],\\
0=&-\frac{1}{2}\alpha_4 \nabla^4 \psi_1 \\
&-\gamma_1\left[\epsilon_{ij}(\partial_i \theta_0)\partial_j -\frac{1}{2}\nabla^2\right]\left[\partial_x \theta_0 -\epsilon_{kl}(\partial_k \theta_0)(\partial_l \psi_1)-\frac{1}{2}\nabla^2 \psi_1\right] .\nonumber
\end{align}
To simplify these equations, we insert Eq.~(\ref{unperturbedtheta}) for $\theta_0$, and change variables to polar coordinates $(r,\phi)$.  We then write $\theta_1(r,\phi)=\theta_r(r)\sin\phi$ and $\psi_1(r,\phi)=\psi_r(r)\sin\phi$.  After those transformations, the differential equations take the form
\begin{align}
0=&K\left[\theta_r''(r)+\frac{\theta_r'(r)}{r}-\frac{\theta_r(r)}{r^2}\right]\nonumber\\
&+\gamma_1 \left[-\frac{\psi_r''(r)}{2}-\frac{(1-2k)\psi_r'(r)}{2r}+\frac{\psi_r(r)}{2r^2}-\frac{k}{r}\right],
\label{thetarequation}\\
0=&\alpha_4 \left[-\frac{\psi_r''''(r)}{2}-\frac{\psi_r'''(r)}{r}+\frac{3\psi_r''(r)}{2r^2}-\frac{3\psi_r'(r)}{2r^3}+\frac{3\psi_r(r)}{2r^4}\right]\nonumber\\
&+\gamma_1\biggl[-\frac{\psi_r''''(r)}{4}-\frac{\psi_r'''(r)}{2r}+\frac{(3-4k+4k^2)\psi_r''(r)}{4r^2}\nonumber\\
&\qquad\quad-\frac{(3-4k+4k^2)\psi_r'(r)}{4r^3}+\frac{(3-4k)\psi_r(r)}{4r^4}+\frac{k^2}{r^3}\biggr].
\label{psirequation}
\end{align}
The solution of Eq.~(\ref{psirequation}) is
\begin{equation}
\psi_r(r)=r+\sum_{i=1}^4 C_i r^{p_i},
\end{equation}
where the exponents $p_i$ are the four roots of the characteristic equation
\begin{align}
0=&\alpha_4 \left[-\frac{p^4}{2}+2p^3-p^2-2p+\frac{3}{2}\right]\\
&+\gamma_1 \left[-\frac{p^4}{4}+p^3-\left(\frac{1}{2}+k-k^2\right)p^2-(1-2k+2k^2)p+\left(\frac{3}{4}-k\right)\right].\nonumber
\end{align}
These roots are
\begin{align}
p=1\pm\biggl[\frac{2}{2+g}&\biggl[2+g(1-k+k^2)\\
&\pm\left[4+2g(2-2k+k^2)+g^2(1-k)^2(1+k^2)\right]^{1/2}\biggr]\biggr]^{1/2},\nonumber
\end{align}
where $g=\gamma_1/\alpha_4$ is the ratio of viscosities.  In general, two of the roots (with $+$ in the first position) are greater than 1, and two of the roots (with $-$ in the first position) are less than 1.

The coefficients $C_i$ are fixed by the boundary conditions.  At the defect, as $r\to0$, we require that the velocity field $\mathbf{v}$ must not diverge, and hence that $\psi$ cannot depend on $r$ with an exponent less than one.  This boundary condition implies that two of the coefficients are zero.  Far from the defect, at a cutoff length $r_\text{max}$, we require that $\psi_r(r_\text{max})=0$ and $\psi_r'(r_\text{max})=0$, so that the velocity field $\mathbf{v}$ also goes to zero.  Those boundary conditions determine the other two coefficients.  Hence, the solution for $\psi_r(r)$ becomes
\begin{equation}
\psi_r(r)=r+\frac{(p_2-1)r_\text{max}^{1-p_1}r^{p_1}}{p_1-p_2}+\frac{(p_1-1)r_\text{max}^{1-p_2}r^{p_2}}{p_2-p_1},
\label{psiresult}
\end{equation}
where $p_1$ and $p_2$ are the two roots with $+$ in the first position.  From that solution, the full stream function becomes $\psi=u\psi_r(r)\sin\phi$, and the flow velocity field becomes $v_i=\epsilon_{ij}\partial_j\psi$.  One interesting consequence of this result is that the velocity field at the defect is $\mathbf{v}(\mathbf{r}\to0)=u\hat{\mathbf{x}}$, which is equal to the velocity of the defect. regardless of the topological charge $k$ and viscosity ratio $g$.  Hence, the fluid flow velocity matches the defect velocity as a result of the calculation, not as a boundary condition.  If the fluid viscosity $\alpha_4$ becomes very high, then the fluid flow velocity decreases very sharply going away from the defect, but it still matches the defect velocity right at the defect core.

To obtain the first-order correction to the director field, we insert the solution for $\psi_r(r)$ into Eq.~(\ref{thetarequation}) and solve for $\theta_r(r)$.  For a boundary conditions, we require that $\theta_r(0)$ does not diverge, and $\theta_r'(r_\text{max})=0$.  The solution is
\begin{align}
\theta_r(r)=\frac{\gamma_1}{2K}\Biggl[&\left(1+\frac{2k(p_1+p_2-p_1 p_2-p_1^2 p_2^2)}{(p_1^2-1)(p_2^2-1)}\right)r
\nonumber\\
&+\frac{(p_1^2-2kp_1-1)(p_2-1)r_\text{max}^{1-p_1}r^{p_1}}{(p_1^2-1)(p_1-p_2)}\nonumber\\
&+\frac{(p_2^2-2kp_2-1)(p_1-1)r_\text{max}^{1-p_2}r^{p_2}}{(p_2^2-1)(p_2-p_1)}\Biggr].
\label{thetaresult}
\end{align}
The full perturbation series for the director field then becomes $\theta=k\phi+\Theta_0 +u\theta_r(r)\sin\phi$.

\begin{figure}
\includegraphics[width=.49\columnwidth]{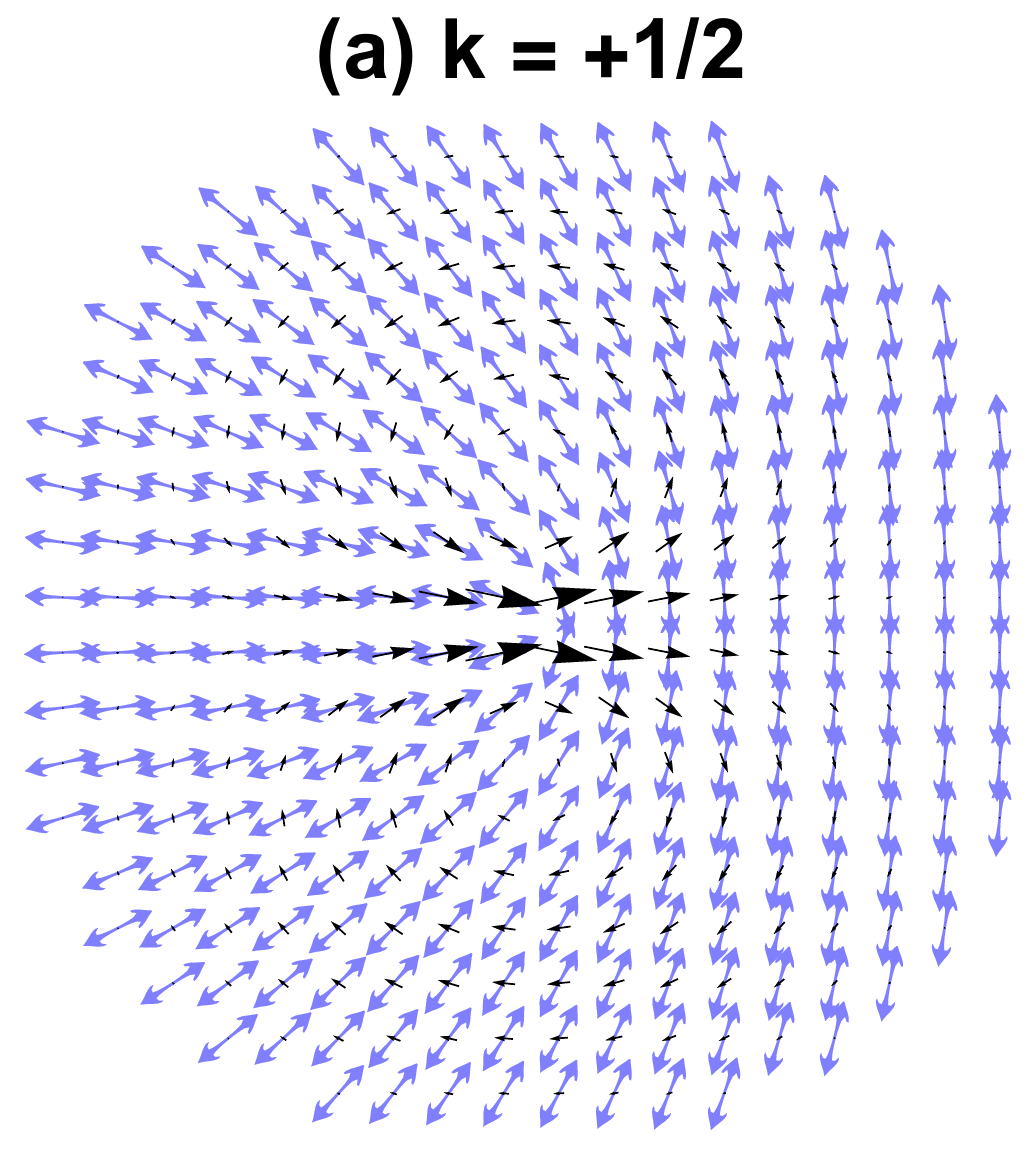}\hfill\includegraphics[width=.49\columnwidth]{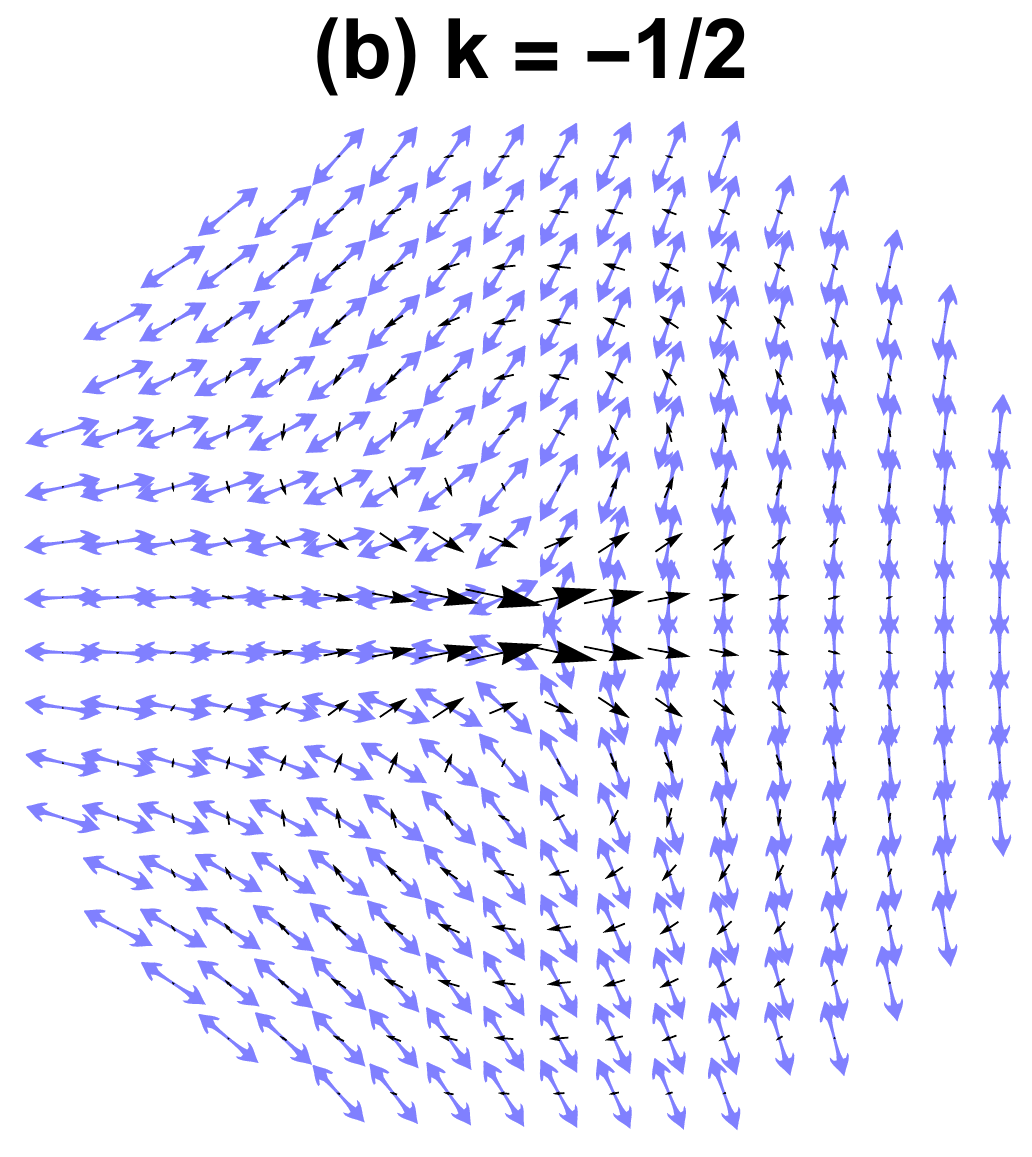}
\caption{Visualization of the results of Eqs.~(\ref{psiresult}) and~(\ref{thetaresult}) for a defect of topological charge $k=\pm1/2$ moving to the right.  Blue double-headed arrows show the director field, and black single-headed arrows show the flow velocity field.  Parameters are $\Theta_0=\pi/2$, $K=1$, $\gamma_1=\alpha_4=1$, $u=1$, and $r_\text{max}=1$.}
\end{figure}

Figure~2 presents examples of the director field and flow velocity field that come from these calculations, for defects of topological charge $k=\pm1/2$ moving to the right.  The director field is slightly distorted compared with the standard arctangent form for the static director field around a topological defect (in a liquid crystal with equal Frank constants).  Because of the factor of $\sin\phi$, the distortion goes to zero in front of and behind the moving defect, and it is greatest in the direction perpendicular to the defect velocity.  This distortion is similar to a recent result for a moving defect in a material that cannot flow.\cite{Radzihovsky2015}  The flow velocity field is greatest at the defect core, and decreases moving away from the defect.  It has a vortex on each side of the moving defect.

We now insert the perturbation series results for $\theta$ and $\psi$ back into Eq.~(\ref{minimalDintermsofthetaandpsi}), to calculate the dissipation function $D$ as a perturbation series in defect velocity $u$.  We integrate from the minimum radius $r_\text{core}$ out to the maximum radius $r_\text{max}$.  The exact integral is quite complicated, and we cannot reproduce it here.  However, it takes a simple and interesting form in the limit of $g=\gamma_1/\alpha_4 \ll 1$, i.e.\ in the limit of high flow viscosity $\alpha_4$, so that the material can only flow very slowly.  When we expand in powers of $g$, the integrated dissipation becomes
\begin{align}
D=&\frac{\pi\gamma_1 k^2 u^2}{2}\log\frac{r_\text{max}}{r_\text{core}}
-\frac{\pi\gamma_1^{3/2}k^2 u^2}{2^{3/2}\alpha_4^{1/2}}\Biggl[|k|\left(\log\frac{r_\text{max}}{r_\text{core}}\right)^2 \nonumber\\
&+\left(2-3k\right)\sign(k)\left(\log\frac{r_\text{max}}{r_\text{core}}\right)-\left(2-\frac{3k}{2}\right)\sign(k)\Biggr].
\label{minimalDresult}
\end{align}
In this expression, the first term is a classic result for the dissipation of a moving defect in a material that cannot flow.\cite{Imura1973}  The second term is a correction to the dissipation in a material that can flow slowly.

Two features of this expression are particularly important.  First, the correction is \emph{negative:}  As the viscosity $\alpha_4$ decreases from infinity to finite values, the drag on a moving defect also decreases.  That result is reasonable, because backflow can partially compensate for the motion of the defect and reduce dissipation.  Second, the classic term is even in topological charge $k$, but the correction term is not.  In a material that cannot flow, with $\alpha_4\to\infty$, there is a symmetry between positive and negative topological charges, which generate equal amounts of dissipation.  In a material that can flow slowly, this symmetry is broken.  The flow pattern reduces the dissipation of positive topological charges more than it reduces the dissipation of negative topological charges.  As a result, negative defects generate more dissipation than positive defects, and hence negative defects will move more slowly than positive defects under the same force.  This flow-induced asymmetry between positive and negative defects has been seen through experiments,\cite{Oswald2005,Blanc2005,Dierking2012} simulations,\cite{Toth2002,Svensek2003} and other theoretical techniques.\cite{Kats2002,Sonnet2005,Sonnet2009}  Here, we see the asymmetry emerge as a specific term of the series expansion in the viscosity ratio $g$.

In Eq.~(\ref{minimalDresult}), the dissipation depends logarithmically on $r_\text{max}$.  The length scale $r_\text{max}$ enters the calculation as a hard cutoff on the dissipation integral, just as it also enters the calculation for the energy of a topological defect.  In a typical experiment with multiple defects, the effective length scale $r_\text{max}$ is given by the characteristic distance between defects.

Some theoretical studies\cite{Pismen1990,Ryskin1991,Denniston1996,Radzihovsky2015} have criticized the dependence on $r_\text{max}$.  Their argument is essentially as follows:  If a defect moves with finite velocity $u$, then the ratio $K/(\gamma_1 u)$ provides a new length scale for the problem, and the dissipation drops off for distances beyond that length scale.  As a result, the dissipation integral really extends out to $r_\text{max}$ or $K/(\gamma_1 u)$, whichever is smaller.  Hence, the result for the dissipation should involve $\log(r_\text{max}/r_\text{core})$ or $\log[K/(\gamma_1 u r_\text{core})]$, whichever is smaller.  If the system is truly infinite, with $r_\text{max}\to\infty$, then $K/(\gamma_1 u)<r_\text{max}$ for any nonzero velocity $u$.  Hence, these studies argue that the dissipation is really proportional to $\log[K/(\gamma_1 u r_\text{core})]$.  Sometimes this dependence is written in terms of the Ericksen number as $\log(3.6/Er)$.  This dissipation can be considered as ``anomalous'' because the dependence on $u$ is not proportional to $u^2$ for small $u$.

Our response to that argument is that it only applies to a system that is strictly infinite.  For any finite system size, there is a crossover velocity $u_c=K/(\gamma_1 r_\text{max})$ at which the cutoff length scale changes.  For $u<u_c$, the dissipation is proportional to $\log(r_\text{max}/r_\text{core})$; for $u>u_c$, it is proportional to $\log[K/(\gamma_1 u r_\text{core})]$.  In general, we want to calculate the dissipation for small $u$ in a finite system, and it scales in the standard way proportional to $u^2$, with a coefficient proportional to $\log(r_\text{max}/r_\text{core})$.  Indeed, this regime is reasonable based on experimental parameters.  In an experiment on defect annihilation,\cite{Dierking2012} the characteristic defect velocity is $u\sim0.3$~$\mu$m/s, and the characteristic distance between defects is $r_\text{max}\sim100$~$\mu$m.  If we assume the Frank constant $K~\sim10^{-11}$~N and rotational viscosity $\gamma_1\sim10^{-1}$~Pa~s, then the crossover velocity is $u_c\sim1$~$\mu$m/s, and we are roughly in the regime of $u<u_c$.

Now we can do the key coarse-graining step:  We compare the dissipation of Eq.~(\ref{minimalDresult}) with the dissipation that would be expected through the macroscopic theory presented in Sec.~2.1.  In this way, we can determine the coefficients of the macroscopic theory.

For a defect of topological charge $k=+1/2$, moving at velocity $\mathbf{u}=u\hat{\mathbf{x}}$ with fixed orientation $\mathbf{p}=(\cos\Psi,\sin\Psi)$, the macroscopic theory of Eq.~(\ref{macroscopicdplushalf}) implies that the dissipation function is
\begin{equation}
D=\frac{1}{2} D_1 |\mathbf{\dot{R}}|^2 + \frac{1}{2} D_2 (\mathbf{p}\cdot\mathbf{\dot{R}})^2
=\frac{1}{2} D_1 u^2 + \frac{1}{2} D_2 u^2 \cos^2 \Psi.
\label{macroscopicdissipationplushalftocompare}
\end{equation}
By comparing Eq.~(\ref{macroscopicdissipationplushalftocompare}) with Eq.~(\ref{minimalDresult}) for $k=+1/2$, we see that the coefficient $D_1$ is
\begin{equation}
D_1 = \frac{\pi\gamma_1}{4}\log\frac{r_\text{max}}{r_\text{core}}
-\frac{\pi\gamma_1^{3/2}}{2^{7/2}\alpha_4^{1/2}}\left[\left(\log\frac{r_\text{max}}{r_\text{core}}\right)^2 
+\left(\log\frac{r_\text{max}}{r_\text{core}}\right)-\frac{5}{2}\right].
\label{minimalD1}
\end{equation}
Furthermore, we see that Eq.~(\ref{minimalDresult}) does not depend on the defect orientation $\Psi$ at all, and hence $D_2=0$.  This lack of dependence on the defect orientation arises because of our minimal model with only the two viscosity coefficients $\alpha_4$ and $\gamma_1$.  In the next section, we will discuss corrections arising from other viscosity coefficients.

For a defect of topological charge $k=-1/2$, again moving at velocity $\mathbf{u}=u\hat{\mathbf{x}}$ with fixed orientation $T_{ijk}$, the the macroscopic theory of Eq.~(\ref{macroscopicdminushalf}) implies the dissipation function
\begin{equation}
D=\frac{1}{2} D'_1 |\mathbf{\dot{R}}|^2 =\frac{1}{2} D_1 u^2 .
\label{macroscopicdissipationminushalftocompare}
\end{equation}
Comparing Eq.~(\ref{macroscopicdissipationminushalftocompare}) with Eq.~(\ref{minimalDresult}) for $k=-1/2$, we find that $D'_1$ is
\begin{equation}
D'_1=\frac{\pi\gamma_1}{4}\log\frac{r_\text{max}}{r_\text{core}}
-\frac{\pi\gamma_1^{3/2}}{2^{7/2}\alpha_4^{1/2}}\left[\left(\log\frac{r_\text{max}}{r_\text{core}}\right)^2 -7\left(\log\frac{r_\text{max}}{r_\text{core}}\right)+\frac{11}{2}\right].
\label{minimalD1prime}
\end{equation}
Here, we see explicitly $D'_1$ and $D_1$ have the same value in the limit of no flow $\alpha_4\to\infty$, but backflow effects reduce $D'_1$ less than they reduce $D_1$.  Thus, we obtain $D'_1>D_1$ in a system with backflow; i.e.\ a negative defect experiences more dissipation and hence more drag.

In addition to the drag coefficients for motion with fixed orientation, the macroscopic theory \emph{also} includes drag coefficients for defect rotation ($D_3$ for a $+1/2$ defect, $D'_3$ for a $-1/2$ defect) and for simultaneous motion and rotation ($D_4$ for a $+1/2$ defect).  Ideally, we would like to calculate those drag coefficients from the same minimal model of liquid-crystal hydrodynamics.  We do not yet have a method for this calculation, because defect rotation is a very long-range distortion that does not decrease with distance from the defect, and hence the dissipation depends sensitively on the boundary conditions.  However, we can at least estimate these coefficients from dimensional analysis.  Because $\mathbf{\dot{p}}$ has one fewer power of length than $\mathbf{\dot{R}}$ in Eq.~(\ref{macroscopicdplushalf}), and $\dot{T}_{ijk}$ has one fewer power of length than $\mathbf{\dot{R}}$ in Eq.~(\ref{macroscopicdminushalf}), we obtain
\begin{align}
D_3 &\sim D'_3 \sim \gamma_1 r_\text{max}^2,\\
D_4 &\sim \gamma_1 r_\text{max}.
\end{align}
Hence, these coefficients diverge with system size $r_\text{max}$ much more severely than do $D_1$ and $D'_1$.  This divergence will be discussed in Sec.~4.

\subsection{Other viscosity coefficients}

The previous section presented a minimal model with only two viscosity coefficients.  However, other viscosity coefficients are also allowed by symmetry in liquid-crystal hydrodynamics.  As noted above, the $(\alpha_5+\alpha_6)$ term is equivalent to the $\alpha_4$ term in 2D (unlike 3D), so we do not need to consider that term separately.  We would like to see how the $\gamma_2$ and $\alpha_1$ coefficients change the macroscopic theory.  

To estimate the effects of those coefficients, we regard the extra terms in the full dissipation function, proportional to $\gamma_2$ and $\alpha_1$, as corrections to the minimal model.  We suppose that $\gamma_2$ and $\alpha_4$ are much smaller than $\gamma_1$ and $\alpha_4$, and calculate these two terms using the director field $\theta(\mathbf{r})=k\phi+\Theta_0 +u\theta_r(r)\sin\phi$ and the flow velocity field $v_i=\epsilon_{ij}\partial_j\psi$, with $\psi(\mathbf{r})=u\psi_r(r)\sin\phi$, which were found from the minimal model in Sec.~2.2.  We do not go back and recalculate the director and flow velocity fields with the other viscosity coefficients.  This procedure is analogous to perturbation theory in quantum mechanics:  At lowest order in perturbation theory, one calculates the expectation value of the perturbed Hamiltonian using the unperturbed wavefunction.  The perturbed wavefunction only enters at higher order.

The term proportional to $\gamma_2$ makes a contribution of
\begin{align}
\int d^2 r \left[\gamma_2 N_i A_{ij} n_j\right] = 0,
\end{align}
except for the special cases of $k=+1$ or $+2$, which we do not discuss here.  Hence, $\gamma_2$ does not affect the drag coefficients, at this order of perturbation theory.  (It may have effects at higher order in perturbation theory.)

By comparison, the term proportional to $\alpha_1$ makes a contribution of
\begin{equation}
\int d^2 r \left[\frac{1}{2}\alpha_1 (n_i A_{ij}n_j)^2\right]
=\frac{\pi\gamma_1 \alpha_1 k^2 u^2}{8\alpha_4}
\left[\left(\log\frac{r_\text{max}}{r_\text{core}}\right)-\frac{1}{2}\right],
\end{equation}
except in the special cases of $k=+1/2$, $+1$, or $+3/2$.  In particular, for $k=+1/2$, we obtain
\begin{align}
\int d^2 r \left[\frac{1}{2}\alpha_1 (n_i A_{ij}n_j)^2\right]
=&\frac{\pi\gamma_1 \alpha_1 u^2}{32\alpha_4}
\left[\left(\log\frac{r_\text{max}}{r_\text{core}}\right)-\frac{1}{2}\right]\\
&+\frac{\pi\gamma_1 \alpha_1 u^2}{64\alpha_4}
\left[\left(\log\frac{r_\text{max}}{r_\text{core}}\right)-1\right]\cos4\Theta_0.\nonumber
\end{align}
The last term is particularly interesting, because it is proportional to $\cos4\Theta_0$.  As discussed earlier, previous papers\cite{Vromans2016,Tang2017} have shown that $\Theta_0$ is related to the defect orientation $\Psi$ by $\Psi=2\Theta_0$, in the case of $+1/2$ defect.  Hence, the last term in the dissipation depends on defect orientation as $\cos2\Psi$, or equivalently as $\cos^2 \Psi$.  This is exactly the orientational dependence that would be expected from the macroscopic drag coefficient $D_2$.  Hence, this orientation-dependent drag coefficient really is present, with a magnitude that scales with the viscosity $\alpha_1$.

\subsection{Example:  Motion of a $\pm1/2$ defect in a channel}

As a simple example to illustrate the microscopic and macroscopic theories, we consider the motion of a defect in a channel.  Although this example is an idealized construction, it is related to dowser textures, which have been studied experimentally and theoretically.\cite{Pieranski2016a,Pieranski2016b}

In this example, we consider a 2D nematic liquid crystal in a channel, which is infinite in the $x$-direction but finite in the $y$-direction.  On the top and bottom surfaces of the channel, at $y=\pm d/2$, there is strong planar anchoring, so that the director field is constrained to be horizontal.  Between those surfaces, the director field may be uniform in the horizontal direction, or it may rotate through an angle of $\pi$.  Indeed, there may be domains of $x$ where the director is uniform or distorted through $\pi$.  In that case, the interface between a uniform domain and a distorted domain is a defect of topological charge $k=\pm1/2$, as shown in Fig.~3(a).

If the defect does not move, then the director field must satisfy the equation for static equilibrium $0=K\nabla^2 \theta$.  An explicit solution that obeys the boundary conditions is
\begin{equation}
\theta(x,y)=\pm\left[\frac{1}{2}\tan^{-1}\left(\frac{\tan(\pi y/d)}{\tanh(\pi x/d)}\right)+\frac{\pi y}{2d}+\frac{\pi}{2}\right],
\label{thetachannel}
\end{equation}
for $k=\pm1/2$.  In general, however, the defect will move in order to reduce the Frank elastic free energy.  In the uniform domain to the left of the defect, the elastic free energy density is 0, but in the distorted domain to the right, it is $\frac{1}{2}K(\pi/d)^2$.  Hence, the defect will move to the right so that the uniform domain will grow, the distorted domain will shrink, and the elastic free energy will decrease.  We can then ask:  What is the velocity of the defect?

In the macroscopic theory, this problem is quite straightforward, and is analogous to the terminal velocity of a particle falling under gravity.  In steady state, the total force acting on the defect must be zero, so that
\begin{equation}
0=\mathbf{f}_\text{elastic}+\mathbf{f}_\text{drag}=-\frac{\partial F}{\partial\mathbf{R}}-\frac{\partial D}{\partial\mathbf{u}}.
\end{equation}
The elastic force is
\begin{equation}
\mathbf{f}_\text{elastic}=-\frac{\partial F}{\partial\mathbf{R}}=\frac{\pi^2 K}{2d}\hat{\mathbf{x}},
\end{equation}
because the elastic free energy decreases by $\frac{1}{2}K(\pi/d)^2 d \delta x$ whenever the defect moves to the right by $\delta x$.  For the $+1/2$ defect, the orientation vector is $\mathbf{p}=-\hat{\mathbf{x}}$, and hence Eq.~(\ref{macroscopicdplushalf}) gives the drag force
\begin{equation}
\mathbf{f}_\text{drag}=-\frac{\partial D}{\partial\mathbf{u}}=-(D_1+D_2)u \hat{\mathbf{x}}.
\end{equation}
Hence, the balance of forces requires that
\begin{equation}
u_{+1/2}=\frac{\pi^2 K}{2d(D_1+D_2)}.
\end{equation}
In the minimal model we have $D_2=0$, and we can estimate $D_1$ by Eq.~(\ref{minimalD1}), with $d/2$ playing the role of $r_\text{max}$.  Hence, the prediction for velocity becomes
\begin{align}
\label{uplus}
u_{+1/2}=&\frac{2\pi K}{\gamma_1 d\log\frac{d}{2r_\text{core}}}\times\\
&\times\left[1+\frac{\gamma_1^{1/2}}{2^{3/2}\alpha_4^{1/2}}
\frac{\left(\log\frac{d}{2r_\text{core}}\right)^2+\left(\log\frac{d}{2r_\text{core}}\right)-\frac{5}{2}}{\log\frac{d}{2r_\text{core}}}+\cdots\right].\nonumber
\end{align}
Similarly, for the $-1/2$ defect, Eq.~(\ref{macroscopicdminushalf}) gives the drag force
\begin{equation}
\mathbf{f}_\text{drag}=-\frac{\partial D}{\partial\mathbf{u}}=-D'_1 u \hat{\mathbf{x}},
\end{equation}
and hence the balance of forces requires that
\begin{equation}
u_{-1/2}=\frac{\pi^2 K}{2d D'_1}.
\end{equation}
In the minimal model, we can estimate $D'_1$ by Eq.~(\ref{minimalD1prime}), with $d/2$ in place of $r_\text{max}$, and hence the prediction for velocity becomes
\begin{align}
\label{uminus}
u_{-1/2}=&\frac{2\pi K}{\gamma_1 d\log\frac{d}{2r_\text{core}}}\times\\
&\times\left[1+\frac{\gamma_1^{1/2}}{2^{3/2}\alpha_4^{1/2}}
\frac{\left(\log\frac{d}{2r_\text{core}}\right)^2-7\left(\log\frac{d}{2r_\text{core}}\right)+\frac{11}{2}}{\log\frac{d}{2r_\text{core}}}+\cdots\right].\nonumber
\end{align}
For a material that cannot flow, with $\alpha_4\to\infty$, the predictions for $u_{+1/2}$ and $u_{-1/2}$ are equal.  As the flow viscosity $\alpha_4$ decreases, both of these velocities increase, but $u_{+1/2}$ increases more than $u_{-1/2}$.  Hence, $+1/2$ defects should move more quickly than $-1/2$ defects because of backflow effects.

To test this macroscopic argument, we perform hydrodynamic simulations of defect motion in a channel.  For these simulations, we must use a formalism based on the 2D nematic order tensor $Q_{ij}=S(2n_i n_j -\delta_{ij})$, so that the scalar order parameter $S$ can go to zero in the defect core.  The minimal model for the free energy is
\begin{equation}
F=-\frac{1}{4}a Q_{ij}Q_{ij}+\frac{1}{16}b(Q_{ij}Q_{ij})^2 + \frac{1}{16}L(\partial_k Q_{ij})(\partial_k Q_{ij}),
\end{equation}
which favors $S=(a/b)^{1/2}$ away from defects.
To represent the rotation of nematic order with respect to the background fluid, instead of the vector $\mathbf{N}$, we use the tensor
\begin{equation}
B_{ij}=\dot{Q}_{ij}-\omega(\epsilon_{lj}Q_{il}+\epsilon_{li}Q_{lj}),
\end{equation}
where again $\omega=\frac{1}{2}\epsilon_{ij}\partial_i v_j$ and $\epsilon_{ij}$ is the 2D Levi-Civita symbol.  The minimal model for the dissipation function then becomes
\begin{equation}
D=\int d^2 r\left[\frac{1}{2}\alpha_4 A_{ij}A_{ij} +\frac{1}{16}\Gamma_1 B_{ij} B_{ij}\right].
\end{equation}
The coefficients $L$ and $\Gamma_1$ in this tensor representation are related to the coefficients $K$ and $\gamma_1$ in the director representation by $K=L S^2$ and $\gamma_1=\Gamma_1 S^2$.

\begin{figure}
\includegraphics[width=\columnwidth]{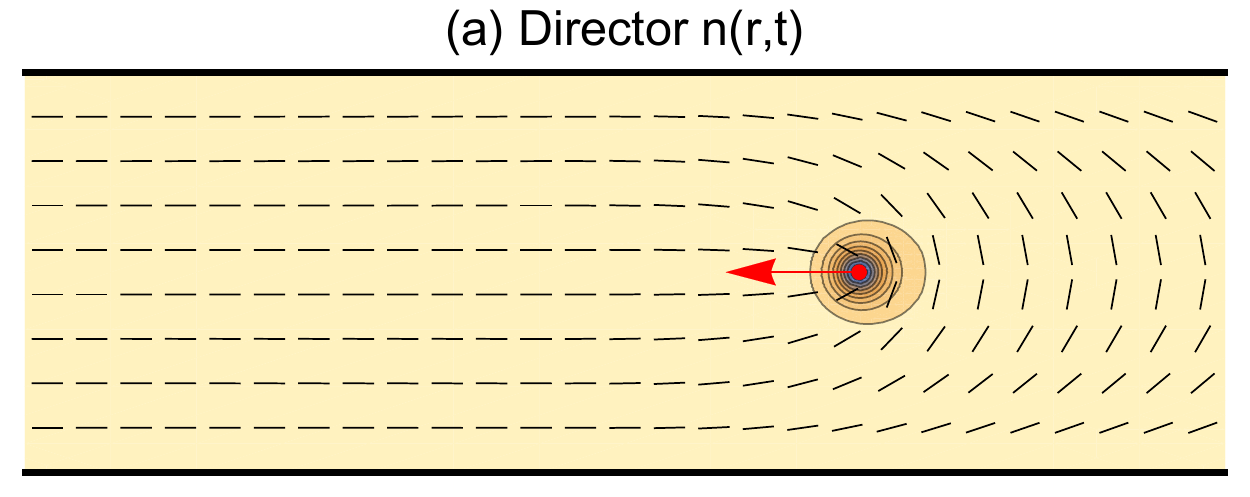}
\includegraphics[width=\columnwidth]{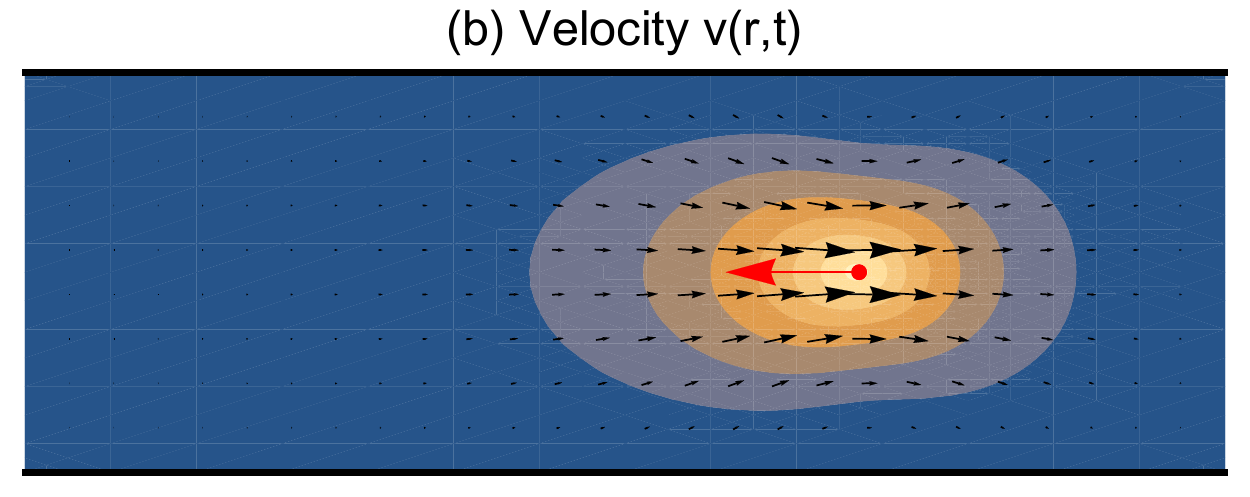}
\caption{Numerical solution of hydrodynamic equations for motion of a $+1/2$ defect toward the right, in a channel with planar boundary conditions.  (a)~Director field shown by black lines, with scalar order parameter indicated by colored contours.  (b)~Velocity field shown by black arrows, with $|\mathbf{v}|^2$ indicated by colored contours.  In both cases, the red arrow represents the defect orientation vector $\mathbf{p}$.  Parameters are $a=b=200$, $L=4$, $\alpha_4 =5$, $\Gamma_1 =8$, $\rho=1$, and $d=2$.  The relatively large values of $a$ and $b$ are chosen to give a relatively small defect core radius $r_\text{core}=(L/a)^{1/2}=0.2$ along with a bulk order parameter $S=(a/b)^{1/2}=1$.}
\end{figure}

We derive partial differential equations for the nematic order tensor $Q_{ij}(\mathbf{r},t)$ and the flow velocity field $v_i(\mathbf{r},t)$ from 
\begin{align}
\label{pdeforQ}
&0=-\frac{\delta F}{\delta Q_{ij}(\mathbf{r},t)}-\frac{\delta D}{\delta[\partial_t Q_{ij}(\mathbf{r},t)]},\\
&\rho\frac{\partial v_i}{\partial t}=-\frac{\delta D}{\delta v_i(\mathbf{r},t)}.
\label{pdeforv}
\end{align}
For computational convenience, we work with constant pressure rather than constant density in this calculation, so that there is no pressure term in the equations, and we use a mass density $\rho$.  We integrate the equations numerically, with planar boundary conditions at $y=\pm d/2$ and open boundary conditions in $x$.  For the initial condition, we use Eq.~(\ref{thetachannel}) for the director orientation around a defect of topological charge $\pm1/2$.  We also assume the initial scalar order parameter drops around the defect core as $S=(a/b)^{1/2}r/(r^2+r_\text{core}^2)^{1/2}$, with a core radius $r_\text{core}=(L/a)^{1/2}$, and the initial flow velocity is zero.

In the numerical solution, the system quickly reaches a steady state, in which the defect moves to the right with constant velocity, and the fluid flow pattern moves along with the defect.  Figure~3 shows an example of the liquid crystal order and fluid flow pattern for a defect of topological charge $+1/2$.  From these numerical results, we can find the defect velocity $u$ for each topological charge as a function of the coefficients $a$, $b$, $L$, $\alpha_4$, and $\Gamma_1$, as well as the channel width $d$.

\begin{figure}
\includegraphics[width=\columnwidth]{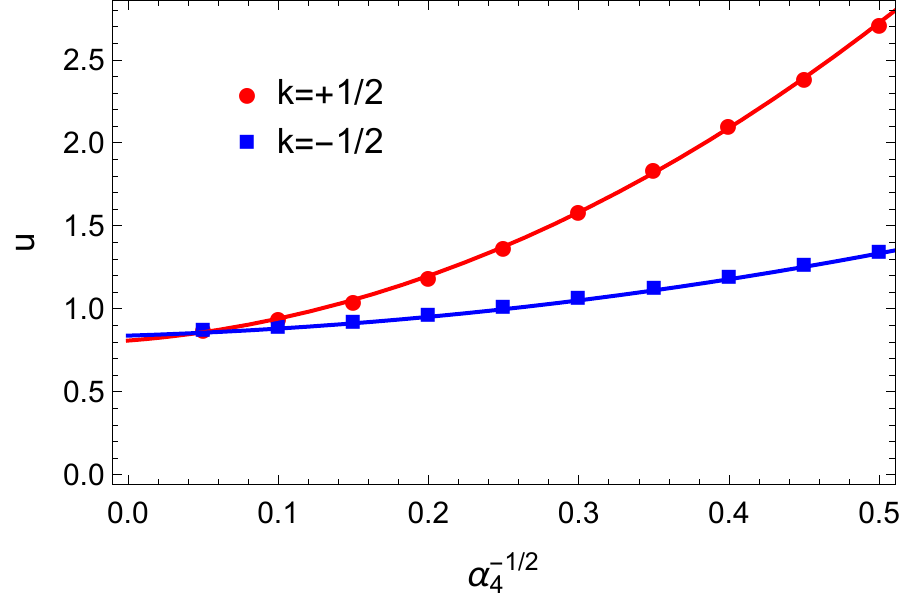}
\caption{Numerical results for the velocities of defects with topological charge $k=\pm1/2$ in the channel geometry, as functions of the flow viscosity $\alpha_4$.  Solid lines show quadratic fits.  Parameters are $a=b=200$, $L=4$, $\Gamma_1 =8$, $\rho=1$, and $d=2$, and hence $r_\text{core}=(L/a)^{1/2}=0.2$ and bulk $S=(a/b)^{1/2}=1$.  Analogous parameters in the director representation are $K=L S^2$ and $\gamma_1=\Gamma_1 S^2$.}
\end{figure}

In Fig.~4, we plot the numerical results for the velocities of $\pm1/2$ defects as functions of fluid flow viscosity, transformed into $\alpha_4^{-1/2}$.  The results are well fit by the quadratic functions
\begin{align}
&u_{+1/2}=0.81+0.69\alpha_4^{-1/2}+6.3\alpha_4^{-1},\nonumber\\
&u_{-1/2}=0.84+0.28\alpha_4^{-1/2}+1.4\alpha_4^{-1}.
\end{align}
We can see that these results are generally consistent with the predictions of the macroscopic theory.  In the limit of high viscosity $\alpha_4\to\infty$, the $\pm1/2$ defects move at approximately the same velocity.  From Eqs.~(\ref{uplus}) and~(\ref{uminus}), we expect that limiting velocity to be $u=2\pi K/[\gamma_1 d \log d/(2r_\text{core})]=0.98$ (with parameters given in the figure caption), which is close to the numerical value.  The numerical velocities may be slightly lower because the hydrodynamic calculation includes extra drag for the motion of the defect core, which is a noticeable fraction of the total area.  As the viscosity $\alpha_4$ decreases, so that the material is able to flow, the velocities for $\pm1/2$ defects both increase, but the velocity for $+1/2$ increases more than the velocity for $-1/2$.  This trend is also consistent with the expectation from Eqs.~(\ref{uplus}) and~(\ref{uminus}), although a quantitative comparison is difficult because those equations are derived assuming $r_\text{max}=\frac{1}{2}d\gg r_\text{core}$, and the hydrodynamic calculation has only a factor of 5 between those values.

Based on this example, we suggest that the macroscopic theory provides a way to develop intuition for the forces that control defect motion in passive liquid crystals.  Furthermore, it gives predictions with far less computational effort than the hydrodynamic approach.  Hence, we would like to extend it to describe active liquid crystals, in which defect motion is even more important.

\section{Active liquid crystals}

\subsection{Coarse-graining the hydrodynamic theory}

In recent years, there has been extensive theoretical and experimental research on active nematic liquid crystals.  These active materials are not in thermal equilibrium, and hence their dynamic behavior is not just driven by minimizing a free energy.  Rather, they continually consume energy, often from a food source or from ATP, and convert this energy into motion.

In the theory of active nematic liquid crystals, the effect of activity is usually modeled by an active contribution to the stress tensor,\cite{Simha2002,Marchetti2013,Giomi2013,Pismen2013,Giomi2014,Prost2015,Zhang2018,Cortese2018,Shankar2018} which can be written in terms of the nematic order tensor as 
\begin{equation}
\sigma^\text{active}_{ij}=-Z Q_{ij} ,
\end{equation}
or in terms of the director as
\begin{equation}
\sigma^\text{active}_{ij}=-\zeta\left(2 n_i n_j -\delta_{ij}\right) .
\end{equation}
Here, the parameter $\zeta=Z S$ is an activity coefficient, with $\zeta>0$ representing a material that tends to \emph{extend} along the director, and $\zeta<0$ indicating a material that tends to \emph{contract} along the director.  This term in the stress tensor contributes to the equation of motion as
\begin{align}
\rho\frac{\partial v_j}{\partial t}&=\text{(passive terms)}+\partial_i\sigma^\text{active}_{ij}\nonumber\\
&=\text{(passive terms)}-2\zeta\partial_i \left(n_i n_j\right).
\label{equationofmotionfromactivestress}
\end{align}
We suggest that the same effect of activity can also be modeled by an active contribution to the dissipation function, which can be written in terms of the nematic order tensor as
\begin{equation}
D^\text{active}=\int d^2 r \left[-Z Q_{ij} A_{ij}\right],
\label{DactivemicroscopicwithQ}
\end{equation}
or in terms of the director as
\begin{equation}
D^\text{active}=\int d^2 r \left[-2\zeta n_i n_j A_{ij}\right],
\label{Dactivemicroscopicwithn}
\end{equation}
with the strain rate tensor $A_{ij}=\frac{1}{2}(\partial_i v_j + \partial_j v_i)$ as before.  This term in the dissipation function contributes to the equation of motion as 
\begin{align}
\rho\frac{\partial v_j}{\partial t}&=\text{(passive terms)}-\frac{\delta D^\text{active}}{\delta v_j}\nonumber\\
&=\text{(passive terms)}-2\zeta\partial_i \left(n_i n_j \right),
\end{align}
which is identical to Eq.~(\ref{equationofmotionfromactivestress}).  Hence, the active term in the dissipation function can be used as a starting point for the theory, equivalent to the active term in the stress tensor.

We recognize that $D^\text{active}$ cannot exactly be regarded as ``energy dissipation,'' because it is not positive-definite.  An alternative description for it might be ``rate of energy input''\cite{Ravnik2018} (with a negative sign).  Nevertheless, it enters into the dissipation function in a formal way, to give the correct equation of motion, so we will use it regardless of the terminology.

We would now like to set up a macroscopic theory for the motion of defects in active nematic liquid crystals.  As in Sec.~2.1, we need to construct the dissipation function in terms of the macroscopic variables that describe a defect.  For a defect of topological charge $k=+1/2$, these variables are the defect position $\mathbf{R}$ and orientation vector $\mathbf{p}$.  At quadratic order in $\mathbf{\dot{R}}$ and $\mathbf{\dot{p}}$, the dissipation function has the passive terms in Eq.~(\ref{macroscopicdplushalf}).  In an active liquid crystal, the dissipation function may include one additional active term that is permitted by symmetry,
\begin{equation}
D^\text{active}=D_5 \mathbf{p}\cdot\mathbf{\dot{R}}.
\label{Dactivemacroscopic}
\end{equation}
This term is not allowed in the dissipation function for a passive liquid crystal because it is odd under time reversal, and hence not positive-definite.  However, it can exist for an active liquid crystal, with the same understanding that is represents rate of energy input (with a negative sign), rather than actual dissipation.

For a defect of topological charge $k=-1/2$, the macroscopic variables are the defect position $\mathbf{R}$ and orientation tensor $T_{ijk}$.  In this case, there is no way to contract the indices to form a nonzero scalar $D^\text{active}$ at linear or quadratic order in velocity $\mathbf{\dot{R}}$ (recalling that $T_{ijk}$ is a completely symmetric tensor with $T_{ijj}=0$).  There could be a cubic term $T_{ijk}\dot{R}_i\dot{R}_j\dot{R}_k$, but it does not affect the motion at low speeds.  Hence, the dynamic behavior of $=-1/2$ defects should be governed by the passive dissipation function of Eq.~(\ref{macroscopicdminushalf}).

In the coarse-graining calculation, we would like to determine how the macroscopic coefficient $D_5$ is related to the more microscopic activity coefficient $\zeta$.  We follow the same procedure as in Sec.~2:  We assume that a defect of topological charge $k=+1/2$ moves with fixed velocity $\mathbf{u}=(u,0)$ at fixed orientation $\mathbf{p}=(\cos\Psi,\sin\Psi)$.  We calculate the dissipation using both microscopic and macroscopic approaches, and compare the results.

For the microscopic calculation, we treat the activity coefficient $\zeta$ in the same way that we treated the viscosity coefficients $(\alpha_5+\alpha_6)$, $\gamma_2$, and $\alpha_1$ in Sec.~2.3:  We regard the active term of Eq.~(\ref{Dactivemicroscopicwithn}) as a perturbation to the minimal model for passive liquid crystals from Sec.~2.2.  Hence, we calculate this term using the director field $\theta(\mathbf{r})=k\phi+\Theta_0 +u\theta_r(r)\sin\phi$ and the flow velocity field $v_i=\epsilon_{ij}\partial_j\psi$, with $\psi(\mathbf{r})=u\psi_r(r)\sin\phi$.  This calculation gives $D^\text{active}=0$ except in the special cases of $k=+1/2$ or $+3/2$.  In particular, for $k=+1/2$, we obtain
\begin{equation}
D^\text{active}=\frac{\pi\zeta\gamma_1^{1/2} u r_\text{max}\cos2\Theta_0}{3 (2\alpha_4)^{1/2}} .
\end{equation}
By comparison, in the macroscopic theory, Eq.~(\ref{Dactivemacroscopic}) implies that $D^\text{active}=D_5 u\cos\Psi$.  Setting these expressions equal, and recalling that $\Psi=2\Theta_0$, we obtain
\begin{equation}
D_5 = \frac{\pi\zeta\gamma_1^{1/2} r_\text{max}}{3 (2\alpha_4)^{1/2}} .
\end{equation}

Several features of this result should be pointed out.  First, it is clearly proportional to the activity coefficient $\zeta$.  It is also proportional to the ratio $(\gamma_1/\alpha_4)^{1/2}$, so that it vanishes in the limit of high fluid flow viscosity $\alpha_4\to\infty$.  That limit is reasonable because the effects of activity require fluid flow.  The result scales linearly with the cutoff length scale $r_\text{max}$, which is a more severe divergence than the logarithmic scaling seen in other terms.

\subsection{Example:  Free motion of a $+1/2$ defect}

As a example, we consider the free motion of a $+1/2$ defect in an active nematic liquid crystal.  In the macroscopic theory, the defect position $\mathbf{R}$ and orientation $\mathbf{p}=(\cos\Psi,\sin\Psi)$ evolve in response to the total forces acting on these macroscopic variables.  If the defect is free, the Frank free energy is constant, and hence there is no elastic force.  Hence, the only forces are the drag forces derived from the dissipation function.  Combining passive and active terms, the full macroscopic dissipation function is
\begin{equation}
D=\frac{1}{2} D_1 |\mathbf{\dot{R}}|^2 + \frac{1}{2} D_2 (\mathbf{p}\cdot\mathbf{\dot{R}})^2 + \frac{1}{2} D_3 |\mathbf{\dot{p}}|^2
+ D_4 \mathbf{\dot{p}}\cdot\mathbf{\dot{R}} + D_5 \mathbf{p}\cdot\mathbf{\dot{R}}.
\label{Dpassiveandactive}
\end{equation}
Hence, the drag force acting on the position is
\begin{equation}
\mathbf{f}_\text{drag}=-\frac{\partial D}{\partial\mathbf{\dot{R}}}
=-D_1\mathbf{\dot{R}}-D_2\mathbf{p}(\mathbf{p}\cdot\mathbf{\dot{R}})-D_4\mathbf{\dot{p}}-D_5 \mathbf{p},
\end{equation}
and the drag force acting on the orientation is
\begin{equation}
f_\text{drag}=-\frac{\partial D}{\partial\dot{\Psi}}=-D_3 \dot{\Psi} -D_4 \mathbf{p}\times\mathbf{\dot{R}}.
\end{equation}

In the steady state, the total force acting on position is zero, and the total force acting on orientation is also zero.  This steady state occurs when
\begin{align}
&\dot{\Psi}=0 \quad\rightarrow\quad \mathbf{\dot{p}}=0,\nonumber\\
&\mathbf{\dot{R}}=-\frac{D_5}{D_1+D_2}\mathbf{p}.
\end{align}
In the minimal model, with $\alpha_4\gg\gamma_1$, this ratio of dissipation coefficients reduces to
\begin{equation}
\mathbf{\dot{R}}=-\frac{4\zeta }{3 (2\gamma_1\alpha_4)^{1/2}}
\frac{r_\text{max}}{\log(r_\text{max}/r_\text{core})}
\mathbf{p}.
\label{predictactivevelocity}
\end{equation}
Hence, in the steady state, the defect moves at a constant velocity with a constant orientation.  The direction of motion is given by the defect orientation $+\mathbf{p}$ if the material is contractile ($\zeta<0$), or $-\mathbf{p}$ if the material is extensile ($\zeta>0$).  The speed is given by the balance between the active force that favors motion and the passive drag force that resists motion.  As a result, the speed is linearly proportional to the activity coefficient $\zeta$ and inversely proportional to the combination of viscosities $(\gamma_1\alpha_4)^{1/2}$.  Also, it is linearly proportional to the cutoff length scale $r_\text{max}$, with a logarithmic correction.  This length scale is generally the system size or the characteristic distance between defects, whichever is smaller.  The linear dependence on cutoff length has been noted in previous work on active liquid crystals.\cite{Giomi2014}

\begin{figure}
\includegraphics[width=\columnwidth]{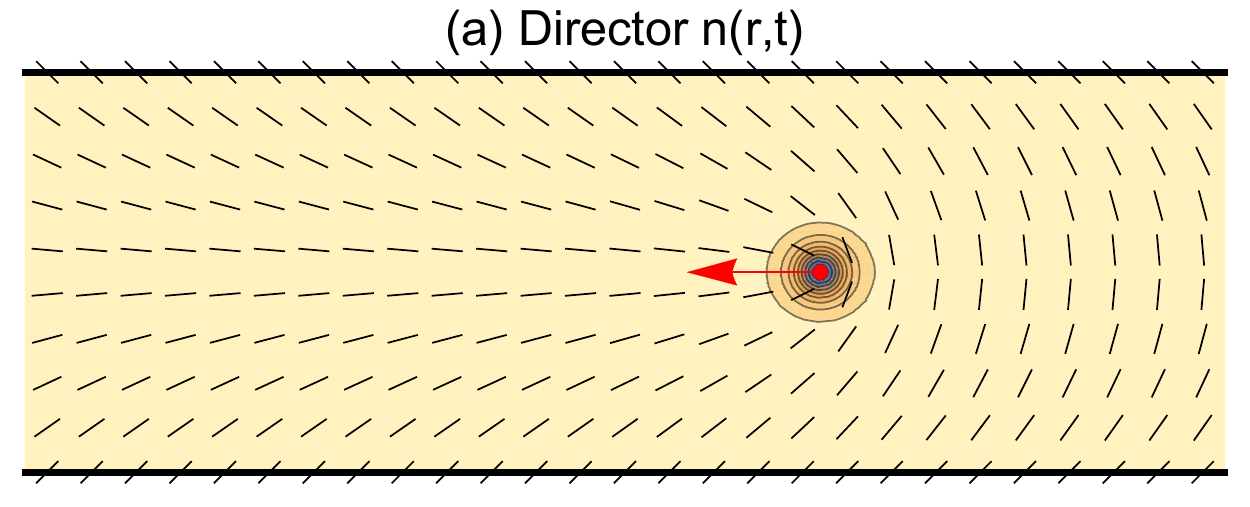}
\includegraphics[width=\columnwidth]{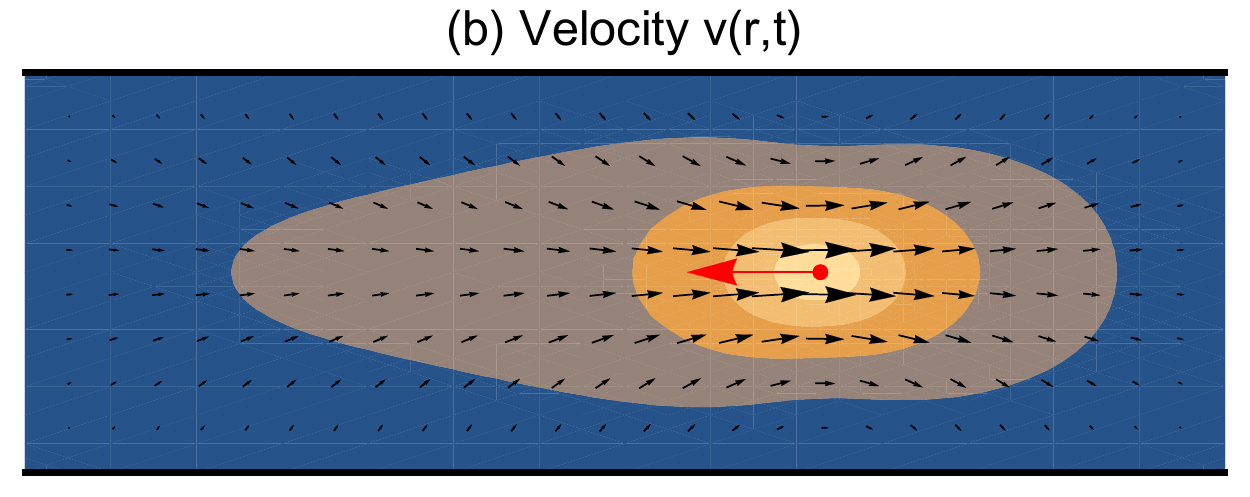}
\caption{Numerical solution of hydrodynamic equations for motion of a $+1/2$ defect toward the right, driven by extensile activity.  (a)~Director field shown by black lines, with scalar order parameter indicated by colored contours.  (b)~Velocity field shown by black arrows, with $|\mathbf{v}|^2$ indicated by colored contours.  In both cases, the red arrow represents the defect orientation vector $\mathbf{p}$.  Parameters are $a=b=200$, $L=4$, $\alpha_4 =1$, $\Gamma_1 =8$, $\rho=1$, $d=2$, and $Z=0.25$.}
\end{figure}

To confirm this macroscopic argument, we perform a modified version of the hydrodynamic simulation for a $+1/2$ defect in Sec.~2.4.  For this modified simulation, we consider a channel with boundary conditions that require the director along the bottom and top surfaces to be at $\theta=\pm\pi/4$, as shown in Fig.~5(a).  Because of that boundary condition, the director must rotate through an angle of $\pi/2$ from bottom to top on both sides of the defect.  Hence, in a system with equal Frank constants, there is equal elastic free energy on both sides of the defect.  As a result, there is no elastic force in the $x$ direction, the defect can move freely in this direction, and the only motion is driven by activity.  Of course, there is still an elastic force that keeps the defect halfway between the walls in the $y$ direction, and an elastic force that keeps the defect orientation at $\Psi=\pi$.

We follow the same method as in Sec.~2.4, using the partial differential equations (\ref{pdeforQ}--\ref{pdeforv}), but with the additional active term of Eq.~(\ref{DactivemicroscopicwithQ}) with coefficient $Z$ in the dissipation function.  The system quickly reaches a steady state, in which the defect moves to the left or right with constant velocity, and the fluid flow pattern moves along with the defect.  Figure~5 shows an example of the liquid crystal order and fluid flow pattern for an extensile material ($Z>0$).  From these simulations, we can find the defect velocity $u$ that is driven by activity.

\begin{figure}
\includegraphics[width=\columnwidth]{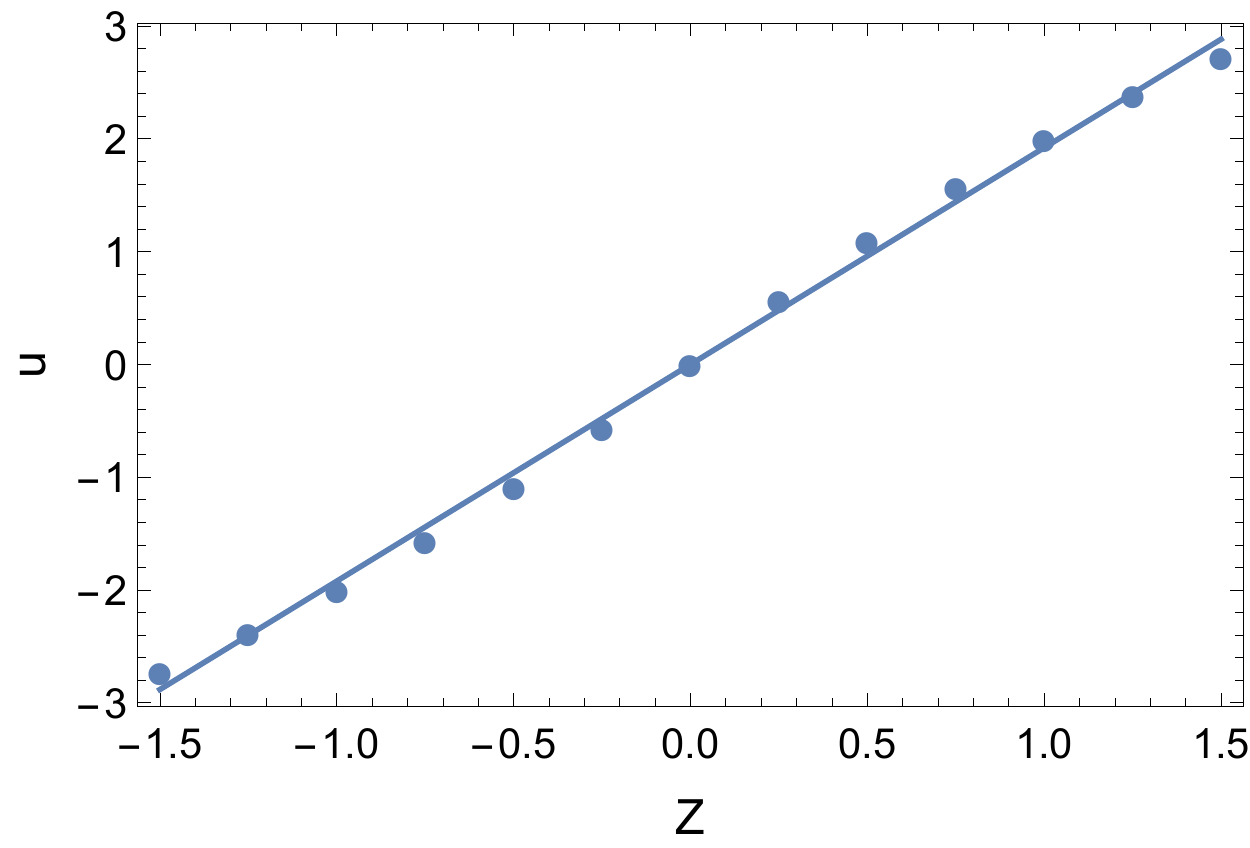}
\caption{Numerical results for the velocity of a $+1/2$ defect as a function of the activity coefficient $Z$.  The solid lines shows a linear fit.  Parameters are $a=b=200$, $L=4$, $\Gamma_1 =8$, $\alpha_4 =1$, $\rho=1$, and $d=2$, and hence $r_\text{core}=(L/a)^{1/2}=0.2$ and bulk $S=(a/b)^{1/2}=1$.  Analogous parameters in the director representation are $K=L S^2$, $\gamma_1=\Gamma_1 S^2$, and $\zeta=Z S$.}
\end{figure}

\begin{figure}
\includegraphics[width=\columnwidth]{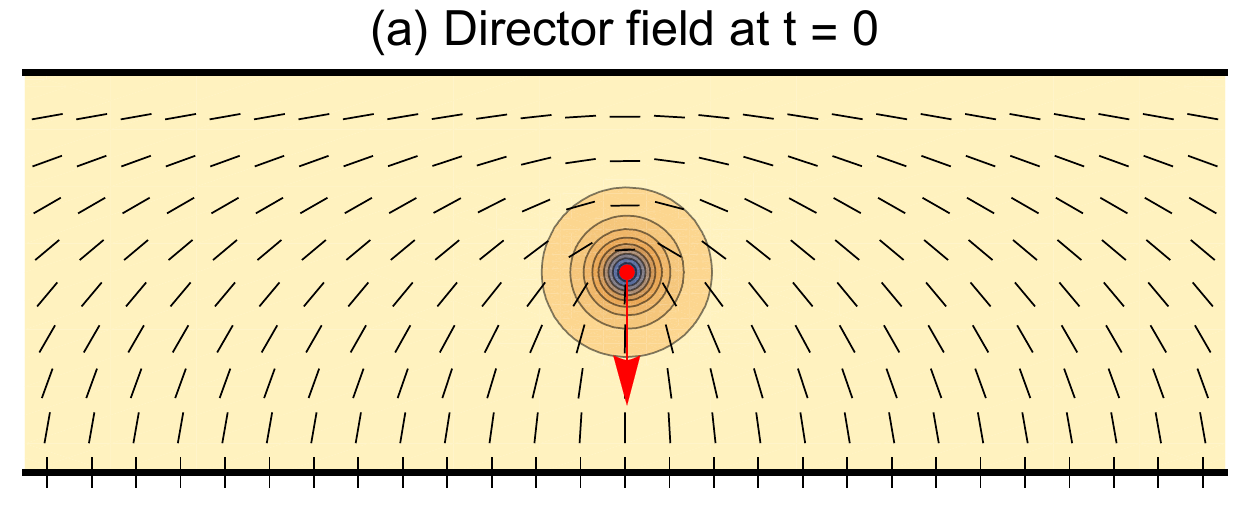}
\includegraphics[width=\columnwidth]{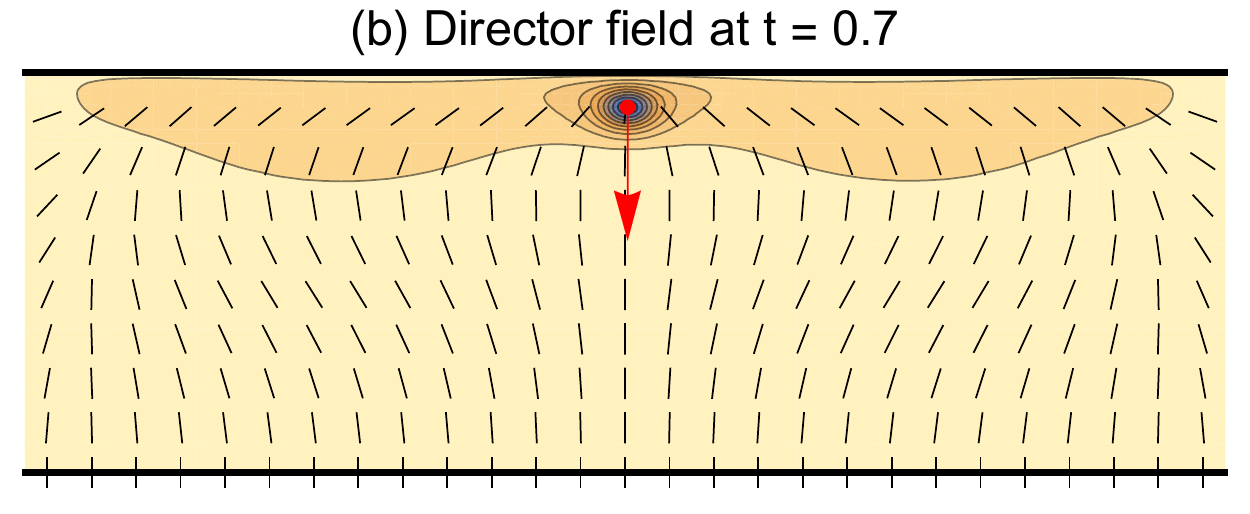}
\includegraphics[width=\columnwidth]{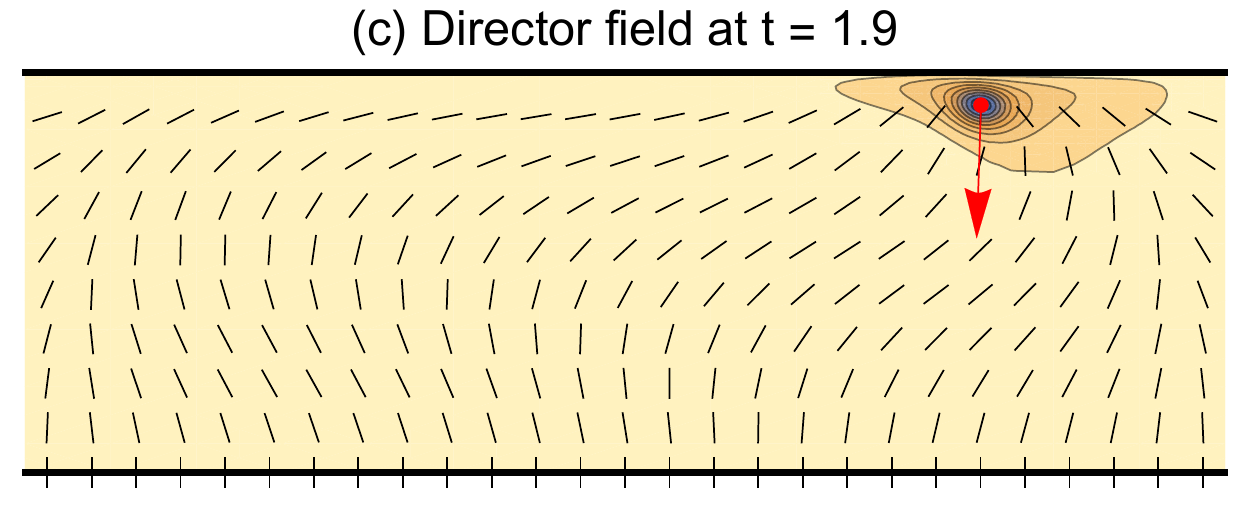}
\caption{Numerical solution of hydrodynamic equations for a $+1/2$ defect pushing against the top wall, driven by extensile activity.  (a)~The defect begins at the center of the channel, with its orientation vector $\mathbf{p}$ pointing downward.  (b)~The defect moves vertically upward until it reaches an equilibrium position at a distance $\delta y$ below the top surface.  (c)~If the activity coefficient $Z$ is greater than a critical value, the orientation $\mathbf{p}$ rotates slightly and the defect moves to the left or right at constant velocity.  In all three visualizations, the director field is shown by black lines, with scalar order parameter indicated by colored contours.  The velocity field is not shown.  Parameters are $a=b=200$, $L=4$, $\alpha_4 =1$, $\Gamma_1 =8$, $\rho=1$, $d=2$, and $Z=30$.}
\end{figure}

In Fig.~6, we plot the numerical results for $u$ as a function of activity coefficient $Z$.  The results are well fit by the straight line $u=1.9 Z$.  By comparison, from Eq.~(\ref{predictactivevelocity}), we expect the relation $u=0.2 Z$ (using $r_\text{max}=\frac{1}{2}d$ and parameters given in the figure caption)).  These relations show the same linear trend, although the quantitative discrepancy in the coefficient indicates a breakdown in some approximation.  One possible issue may be that the analytic calculation was done in a circular geometry of radius $r_\text{max}$, while the simulation was done in a rectangular geometry.  For quantities like the passive drag force, which diverge logarithmically with system size, it is generally reasonable to approximate a rectangle by a circle with radius equal to the smaller rectangular dimension.  This may not be a reasonable approximation for the active driving force, which diverges more severely with system size.

\subsection{Example:  $+1/2$ defect pushing against wall}

For a further example of defect motion, we modify the boundary conditions on the channel so that it requires planar alignment on the top surface and homeotropic alignment on the bottom surface.  As a result, the system can form a defect of topological charge $+1/2$ with orientation vector $\mathbf{p}=-\hat{\mathbf{y}}$, as shown in Fig.~7(a).  When the system evolves with extensile activity $Z>0$, the defect moves vertically toward the top surface, and it pushes against that wall, as in Fig.~7(b).  After that, the behavior depends on the magnitude of the activity.  If the activity is less than a critical value, the defect remains stable while pushing against the wall.  By contrast, if the activity is greater than the critical value, the defect remains approximately stationary for some time, and then eventually breaks the symmetry between the $\pm\hat{\mathbf{x}}$ directions.  At that time, it rotates its orientation slightly, and moves to the left or the right at constant velocity, as in Fig.~7(c).  This symmetry-breaking behavior is similar to the formation of a ``yin-yang'' structure by two $+1/2$ defects pushing outward against a circular wall.\cite{Norton2018}

In the simulation, we use fixed boundary conditions on the left and right edges, and hence the defect bounces off these edges and moves back and forth between left and right.  Presumably, if the simulation were infinite in the $x$-direction, then the horizontal motion would continue at fixed velocity without limit.

This motion can be understood from the macroscopic view of a defect as an oriented particle.  In the macroscopic view, the free energy arises from the interaction of the defect with the top aligning surface, which can equivalently be regarded as the interaction of the defect with an image defect above the top surface.  Following the argument in our previous paper,\cite{Tang2017} this free energy becomes
\begin{equation}
F\approx -K \log\left(\frac{\delta y}{r_\text{core}}\right) + \frac{1}{2}K(\delta\Psi)^2,
\end{equation}
where $K$ is the Frank constant, $\delta y = y_\text{max}-y$ is the distance from the top surface, and $\delta\Psi=\Psi+\pi/2$ is the defect orientation relative to the favored orientation of $-\pi/2$.  The dissipation function is still the same combination of passive and active terms as in Eq.~(\ref{Dpassiveandactive}).

In the first stage of motion, the defect moves upward until it reaches an equilibrium point at a fixed $\delta y$.  At that point, the elastic force pushing downward is $-\partial F/\partial y=-K/\delta y$, and the active force pushing upward is $-\partial u/\partial\dot{y}=D_5$.  Hence, the equilibrium occurs at the position
\begin{equation}
\delta y = \frac{K}{D_5}
= \frac{3K}{\pi\zeta r_\text{max}} \left(\frac{2\alpha_4}{\gamma_1}\right)^{1/2}
= \frac{3L}{\pi Z r_\text{max}} \left(\frac{2\alpha_4}{\Gamma_1}\right)^{1/2}.
\end{equation}
We can assume that the cutoff distance is $r_\text{max}\approx\delta y$, because that is the distance from the defect to the nearest boundary.  Hence, we obtain
\begin{equation}
\delta y \approx \left(\frac{3L}{\pi Z}\right)^{1/2} \left(\frac{2\alpha_4}{\Gamma_1}\right)^{1/4}.
\end{equation}
For the parameters in the simulation, this prediction gives $\delta y = 1.4 Z^{-1/2}$.  In comparison, Fig.~8(a) shows the numerical results for $\delta y$ as a function of $Z$, which are well fit by $\delta y = 1.0 Z^{-1/2}$.  This agreement is reasonably good, considering the roughness of our estimate for the cutoff distance.

\begin{figure}
\includegraphics[width=\columnwidth]{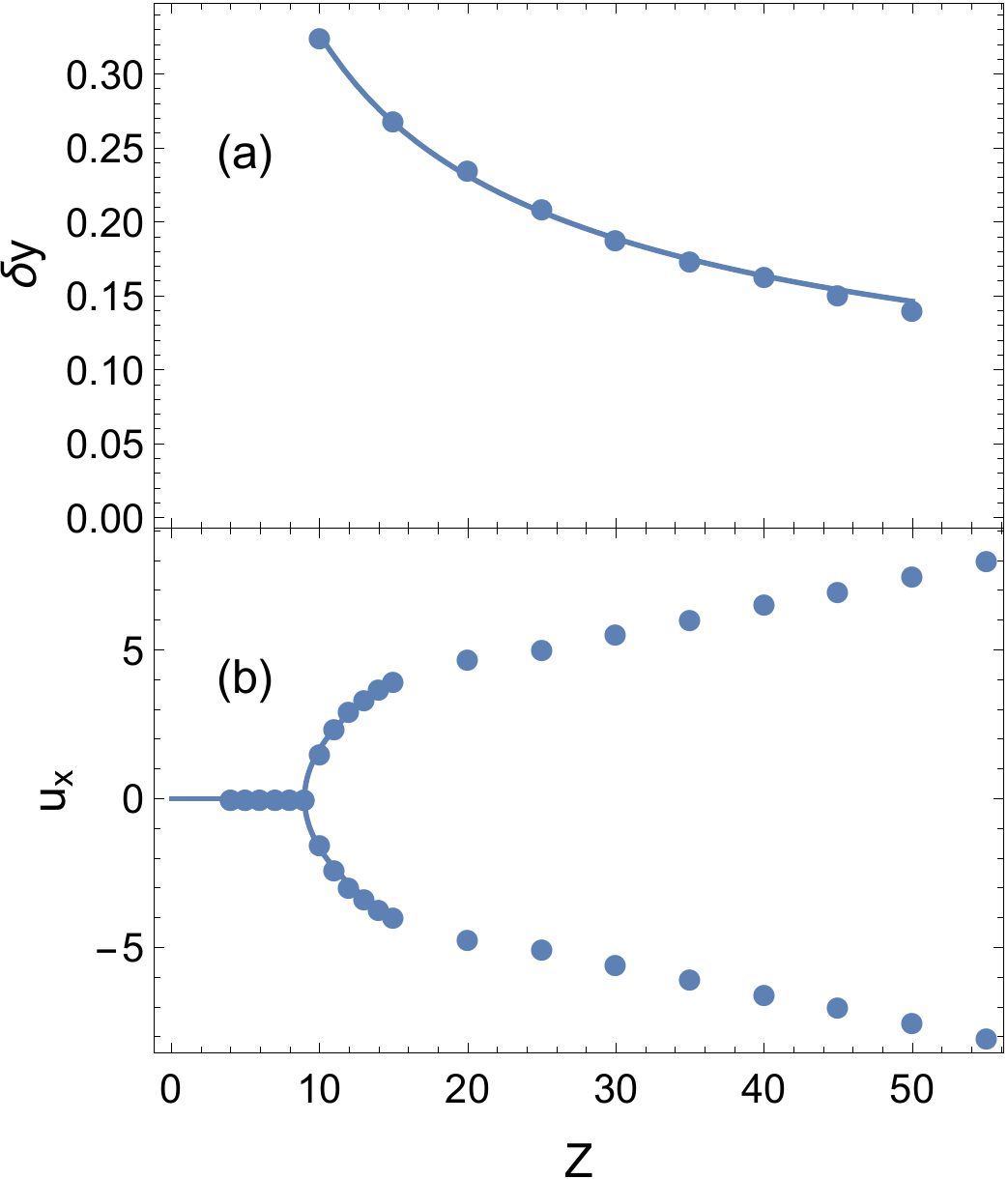}
\caption{Numerical results for a $+1/2$ defect pushing against the top wall, driven by extensile activity.  (a)~Distance $\delta y$ below the top surface, as a function of activity coefficient $Z$, showing the scaling with $Z^{-1/2}$.  (b)~Velocity $u_x$, also as a function of $Z$, showing the symmetry-breaking square-root bifurcation from the stationary state at low activity to the moving state at high activity.  Parameters are $a=b=200$, $L=4$, $\Gamma_1 =8$, $\alpha_4 =1$, $\rho=1$, and $d=2$, and hence $r_\text{core}=(L/a)^{1/2}=0.2$ and bulk $S=(a/b)^{1/2}=1$.  Analogous parameters in the director representation are $K=L S^2$, $\gamma_1=\Gamma_1 S^2$, and $\zeta=Z S$.}
\end{figure}

In the second stage of motion, the defect orientation $\Psi$ rotates slightly, and the defect moves to the right or left with constant $\dot{x}$, $y$, and $\Psi$.  The elastic force on $\Psi$ is $-\partial F/\partial\Psi=-K\delta\Psi$, and the drag force on $\Psi$ is $-\partial D/\partial\dot{\Psi}=-D_4\dot{x}\cos\delta\Psi$.  Likewise, the elastic force on $x$ is zero, and the (passive plus active) drag force on $x$ is $-\partial D/\partial\dot{x}=-D_1\dot{x}-D_5\sin\delta\Psi$ (neglecting the $D_2$ anisotropy).  Balancing these two pairs of forces gives two simultaneous equations in $\delta\Psi$ and $\dot{x}$.  The trivial solution to these equations is $\delta\Psi=\dot{x}=0$.  When the active coefficient $D_5$ exceeds the critical value $D_5^\text{crit}=D_1 K/D_4$, there is a bifurcation to the nontrivial solution
\begin{align}
&\delta\Psi=\mp\left[\frac{3}{2}\frac{D_5-D_5^\text{crit}}{D_5^\text{crit}}\right]^{1/2}
\propto\mp\left[Z-Z^\text{crit}\right]^{1/2},\nonumber\\
&\dot{x}=\pm\frac{1}{D_1}\left[\frac{3}{2}D_5^\text{crit}(D_5-D_5^\text{crit})\right]^{1/2}
\propto\pm\left[Z-Z^\text{crit}\right]^{1/2} .
\end{align}
Hence, we expect a classic square-root bifurcation as a function of activity, beyond a finite critical activity.  For comparison, Fig.~8(b) shows the numerical results for velocity $\dot{x}$ as a function of $Z$.  We can see the bifurcation at $Z^\text{crit}$, with the velocity scaling as a square root for activity just above that point.  Hence, the macroscopic view of defect dynamics effectively describes this instability.

\section{Discussion}

In this paper, we have combined the concept of defects as particles moving under forces with the concept of defect orientation.  In this combined view, defects are effective particles with both position and orientation.  Forces may act to change the position, orientation, or both.  Hence, to predict the motion of defects, we must balance the elastic and drag forces acting on both position and orientation.  For \emph{passive} liquid crystals, the concept of defect orientation is moderately important, because it is an additional macroscopic degree of freedom, which can modify the elastic and drag forces.  For \emph{active} liquid crystals, the concept of defect orientation is even more important, because it defines the direction of the active force.  In particular, $+1/2$ defects move as particles with a vector orientation, in a characteristic direction, while $-1/2$ defects move as particles with three-fold symmetry.

In addition to these specific results about defect orientation, we have explored the general formalism of the Rayleigh dissipation function as an approach to model the dynamics of liquid crystals.  We see that this formalism can describe active dynamics, with activity appearing as a negative contribution to the dissipation.  Of course, this negative contribution is not exactly dissipation; it might better be called the rate of energy input.  Even so, it enters the dissipation function in a formal way to give the same equations of motion that have already been derived from an active stress tensor.  Hence, it allows active forces to be modeled through the same approach as passive drag forces.  We also see that the dissipation function formalism provides a way to coarse-grain the dynamics.  By equating the dissipation functions calculated through different methods, one can go from the hydrodynamic theory of the director and velocity fields to the more macroscopic description of defects as effective oriented particles.

The greatest strength of the macroscopic description of defects as particles is to provide an intuitive understanding of defect motion.  By considering all the forces acting on defect position and orientation, we can see how defects will move in either passive or active liquid crystals.  The usefulness of this approach is demonstrated in the example of Sec.~3.3.  In the macroscopic view of the defect as an oriented particle pushing against a wall, we can easily see that it should be stationary for low activity, but it should tilt and move horizontally for high activity.  By contrast, in the hydrodynamic theory of the director and velocity fields, it is challenging to solve the partial differential equations for time evolution, and the result is not obvious.

By contrast, the greatest weakness of the macroscopic description is that the macroscopic drag coefficients $D_1$ through $D_5$ diverge logarithmically, or even more severely, as the cutoff length scale $r_\text{max}\to\infty$.  These divergences occur because defects create long-range distortions in the nematic director field, which only decay slowly with distance from the defect core.  Because of these divergences, the drag coefficients can change with the defect environment, especially as a defect gets close to a boundary or to other defects.  That change is seen explicitly in the first stage of motion in Sec.~3.3, in the scaling behavior of $\delta y$ with activity.  This dependence on environment makes it more difficult to use the macroscopic approach for quantitative predictions of motion.

We note that 3D liquid crystals exhibit other types of moving structures, which are more localized than the 2D disclination defects studied here.  These structures include skyrmions, topological configurations of a 3D director field, which ``squirm'' under an applied electric field.\cite{Ackerman2017}  They also include bullet-like solitons, which form under an electric field and move rapidly across a sample.\cite{Li2018}  In future work, the macroscopic approach to dynamics might be applied to these structures.  Because the director distortions are localized, we expect that all of the integrals for drag coefficients should converge.  As a result, these structures might be even more effectively described as particles, with drag properties that are less dependent on their environment.

We would like to thank A.~Baskaran for helpful discussions.  This work was supported by National Science Foundation Grant No.~DMR-1409658. 





\bibliography{defectmotion2} 

\providecommand*{\mcitethebibliography}{\thebibliography}
\csname @ifundefined\endcsname{endmcitethebibliography}
{\let\endmcitethebibliography\endthebibliography}{}
\begin{mcitethebibliography}{57}
\providecommand*{\natexlab}[1]{#1}
\providecommand*{\mciteSetBstSublistMode}[1]{}
\providecommand*{\mciteSetBstMaxWidthForm}[2]{}
\providecommand*{\mciteBstWouldAddEndPuncttrue}
  {\def\EndOfBibitem{\unskip.}}
\providecommand*{\mciteBstWouldAddEndPunctfalse}
  {\let\EndOfBibitem\relax}
\providecommand*{\mciteSetBstMidEndSepPunct}[3]{}
\providecommand*{\mciteSetBstSublistLabelBeginEnd}[3]{}
\providecommand*{\EndOfBibitem}{}
\mciteSetBstSublistMode{f}
\mciteSetBstMaxWidthForm{subitem}
{(\emph{\alph{mcitesubitemcount}})}
\mciteSetBstSublistLabelBeginEnd{\mcitemaxwidthsubitemform\space}
{\relax}{\relax}

\bibitem[Chuang \emph{et~al.}({1991})Chuang, Durrer, Turok, and
  Yurke]{Chuang1991}
I.~Chuang, R.~Durrer, N.~Turok and B.~Yurke, \emph{{Science}}, {1991},
  \textbf{{251}}, {1336--1342}\relax
\mciteBstWouldAddEndPuncttrue
\mciteSetBstMidEndSepPunct{\mcitedefaultmidpunct}
{\mcitedefaultendpunct}{\mcitedefaultseppunct}\relax
\EndOfBibitem
\bibitem[Bowick \emph{et~al.}(1994)Bowick, Chandar, Schiff, and
  Srivastava]{Bowick1994}
M.~J. Bowick, L.~Chandar, E.~A. Schiff and A.~M. Srivastava, \emph{Science},
  1994, \textbf{263}, 943--945\relax
\mciteBstWouldAddEndPuncttrue
\mciteSetBstMidEndSepPunct{\mcitedefaultmidpunct}
{\mcitedefaultendpunct}{\mcitedefaultseppunct}\relax
\EndOfBibitem
\bibitem[Pargellis \emph{et~al.}({1991})Pargellis, Turok, and
  Yurke]{Pargellis1991}
A.~Pargellis, N.~Turok and B.~Yurke, \emph{{Phys. Rev. Lett.}}, {1991},
  \textbf{{67}}, {1570--1573}\relax
\mciteBstWouldAddEndPuncttrue
\mciteSetBstMidEndSepPunct{\mcitedefaultmidpunct}
{\mcitedefaultendpunct}{\mcitedefaultseppunct}\relax
\EndOfBibitem
\bibitem[Pargellis \emph{et~al.}({1992})Pargellis, Finn, Goodby, Panizza,
  Yurke, and Cladis]{Pargellis1992}
A.~N. Pargellis, P.~Finn, J.~W. Goodby, P.~Panizza, B.~Yurke and P.~E. Cladis,
  \emph{{Phys. Rev. A}}, {1992}, \textbf{{46}}, {7765--7776}\relax
\mciteBstWouldAddEndPuncttrue
\mciteSetBstMidEndSepPunct{\mcitedefaultmidpunct}
{\mcitedefaultendpunct}{\mcitedefaultseppunct}\relax
\EndOfBibitem
\bibitem[Oswald and Ignes-Mullol({2005})]{Oswald2005}
P.~Oswald and J.~Ignes-Mullol, \emph{{Phys. Rev. Lett.}}, {2005},
  \textbf{{95}}, {027801}\relax
\mciteBstWouldAddEndPuncttrue
\mciteSetBstMidEndSepPunct{\mcitedefaultmidpunct}
{\mcitedefaultendpunct}{\mcitedefaultseppunct}\relax
\EndOfBibitem
\bibitem[Blanc \emph{et~al.}(2005)Blanc, Sven\ifmmode~\check{s}\else
  \v{s}\fi{}ek, \ifmmode~\check{Z}\else \v{Z}\fi{}umer, and Nobili]{Blanc2005}
C.~Blanc, D.~Sven\ifmmode~\check{s}\else \v{s}\fi{}ek,
  S.~\ifmmode~\check{Z}\else \v{Z}\fi{}umer and M.~Nobili, \emph{Phys. Rev.
  Lett.}, 2005, \textbf{95}, 097802\relax
\mciteBstWouldAddEndPuncttrue
\mciteSetBstMidEndSepPunct{\mcitedefaultmidpunct}
{\mcitedefaultendpunct}{\mcitedefaultseppunct}\relax
\EndOfBibitem
\bibitem[Stannarius \emph{et~al.}({2006})Stannarius, Bohley, and
  Eremin]{Stannarius2006}
R.~Stannarius, C.~Bohley and A.~Eremin, \emph{{Phys. Rev. Lett.}}, {2006},
  \textbf{{97}}, {097802}\relax
\mciteBstWouldAddEndPuncttrue
\mciteSetBstMidEndSepPunct{\mcitedefaultmidpunct}
{\mcitedefaultendpunct}{\mcitedefaultseppunct}\relax
\EndOfBibitem
\bibitem[Dierking \emph{et~al.}({2012})Dierking, Ravnik, Lark, Healey,
  Alexander, and Yeomans]{Dierking2012}
I.~Dierking, M.~Ravnik, E.~Lark, J.~Healey, G.~P. Alexander and J.~M. Yeomans,
  \emph{{Phys. Rev. E}}, {2012}, \textbf{{85}}, {021703}\relax
\mciteBstWouldAddEndPuncttrue
\mciteSetBstMidEndSepPunct{\mcitedefaultmidpunct}
{\mcitedefaultendpunct}{\mcitedefaultseppunct}\relax
\EndOfBibitem
\bibitem[Guimaraes \emph{et~al.}({2013})Guimaraes, Mendes, Fernandes, and
  Mukai]{Guimaraes2013}
R.~R. Guimaraes, R.~S. Mendes, P.~R.~G. Fernandes and H.~Mukai, \emph{{J. Phys.
  Condens. Matter}}, {2013}, \textbf{{25}}, {404203}\relax
\mciteBstWouldAddEndPuncttrue
\mciteSetBstMidEndSepPunct{\mcitedefaultmidpunct}
{\mcitedefaultendpunct}{\mcitedefaultseppunct}\relax
\EndOfBibitem
\bibitem[Kim \emph{et~al.}({2013})Kim, Shiyanovskii, and Lavrentovich]{Kim2013}
Y.-K. Kim, S.~V. Shiyanovskii and O.~D. Lavrentovich, \emph{{J. Phys. Condens.
  Matter}}, {2013}, \textbf{{25}}, {404202}\relax
\mciteBstWouldAddEndPuncttrue
\mciteSetBstMidEndSepPunct{\mcitedefaultmidpunct}
{\mcitedefaultendpunct}{\mcitedefaultseppunct}\relax
\EndOfBibitem
\bibitem[Stannarius and Harth({2016})]{Stannarius2016}
R.~Stannarius and K.~Harth, \emph{{Phys. Rev. Lett.}}, {2016}, \textbf{{117}},
  {157801}\relax
\mciteBstWouldAddEndPuncttrue
\mciteSetBstMidEndSepPunct{\mcitedefaultmidpunct}
{\mcitedefaultendpunct}{\mcitedefaultseppunct}\relax
\EndOfBibitem
\bibitem[Sanchez \emph{et~al.}({2012})Sanchez, Chen, DeCamp, Heymann, and
  Dogic]{Sanchez2012}
T.~Sanchez, D.~T.~N. Chen, S.~J. DeCamp, M.~Heymann and Z.~Dogic,
  \emph{{Nature}}, {2012}, \textbf{{491}}, {431--434}\relax
\mciteBstWouldAddEndPuncttrue
\mciteSetBstMidEndSepPunct{\mcitedefaultmidpunct}
{\mcitedefaultendpunct}{\mcitedefaultseppunct}\relax
\EndOfBibitem
\bibitem[Keber \emph{et~al.}({2014})Keber, Loiseau, Sanchez, DeCamp, Giomi,
  Bowick, Marchetti, Dogic, and Bausch]{Keber2014}
F.~C. Keber, E.~Loiseau, T.~Sanchez, S.~J. DeCamp, L.~Giomi, M.~J. Bowick,
  M.~C. Marchetti, Z.~Dogic and A.~R. Bausch, \emph{{Science}}, {2014},
  \textbf{{345}}, {1135--1139}\relax
\mciteBstWouldAddEndPuncttrue
\mciteSetBstMidEndSepPunct{\mcitedefaultmidpunct}
{\mcitedefaultendpunct}{\mcitedefaultseppunct}\relax
\EndOfBibitem
\bibitem[DeCamp \emph{et~al.}({2015})DeCamp, Redner, Baskaran, Hagan, and
  Dogic]{DeCamp2015}
S.~J. DeCamp, G.~S. Redner, A.~Baskaran, M.~F. Hagan and Z.~Dogic, \emph{{Nat.
  Mater.}}, {2015}, \textbf{{14}}, {1110--1115}\relax
\mciteBstWouldAddEndPuncttrue
\mciteSetBstMidEndSepPunct{\mcitedefaultmidpunct}
{\mcitedefaultendpunct}{\mcitedefaultseppunct}\relax
\EndOfBibitem
\bibitem[Ericksen({1960})]{Ericksen1960}
J.~L. Ericksen, \emph{{Arch. Ration. Mech. Anal.}}, {1960}, \textbf{{4}},
  {231--237}\relax
\mciteBstWouldAddEndPuncttrue
\mciteSetBstMidEndSepPunct{\mcitedefaultmidpunct}
{\mcitedefaultendpunct}{\mcitedefaultseppunct}\relax
\EndOfBibitem
\bibitem[Ericksen({1961})]{Ericksen1961}
J.~L. Ericksen, \emph{{Trans. Soc. Rheol.}}, {1961}, \textbf{{5}},
  {23--34}\relax
\mciteBstWouldAddEndPuncttrue
\mciteSetBstMidEndSepPunct{\mcitedefaultmidpunct}
{\mcitedefaultendpunct}{\mcitedefaultseppunct}\relax
\EndOfBibitem
\bibitem[Leslie({1966})]{Leslie1966}
F.~M. Leslie, \emph{{Q. J. Mech. Appl. Math}}, {1966}, \textbf{{19}},
  {357--370}\relax
\mciteBstWouldAddEndPuncttrue
\mciteSetBstMidEndSepPunct{\mcitedefaultmidpunct}
{\mcitedefaultendpunct}{\mcitedefaultseppunct}\relax
\EndOfBibitem
\bibitem[Leslie({1968})]{Leslie1968}
F.~M. Leslie, \emph{{Arch. Ration. Mech. Anal.}}, {1968}, \textbf{{28}},
  {265--283}\relax
\mciteBstWouldAddEndPuncttrue
\mciteSetBstMidEndSepPunct{\mcitedefaultmidpunct}
{\mcitedefaultendpunct}{\mcitedefaultseppunct}\relax
\EndOfBibitem
\bibitem[Beris and Edwards(1994)]{Beris1994}
A.~Beris and B.~Edwards, \emph{Thermodynamics of Flowing Systems}, Oxford,
  1994\relax
\mciteBstWouldAddEndPuncttrue
\mciteSetBstMidEndSepPunct{\mcitedefaultmidpunct}
{\mcitedefaultendpunct}{\mcitedefaultseppunct}\relax
\EndOfBibitem
\bibitem[T{\'{o}}th \emph{et~al.}(2002)T{\'{o}}th, Denniston, and
  Yeomans]{Toth2002}
G.~T{\'{o}}th, C.~Denniston and J.~M. Yeomans, \emph{Phys. Rev. Lett.}, 2002,
  \textbf{88}, 105504\relax
\mciteBstWouldAddEndPuncttrue
\mciteSetBstMidEndSepPunct{\mcitedefaultmidpunct}
{\mcitedefaultendpunct}{\mcitedefaultseppunct}\relax
\EndOfBibitem
\bibitem[Sven{\v{s}}ek and {\v{Z}}umer(2003)]{Svensek2003}
D.~Sven{\v{s}}ek and S.~{\v{Z}}umer, \emph{Phys. Rev. Lett.}, 2003,
  \textbf{90}, 155501\relax
\mciteBstWouldAddEndPuncttrue
\mciteSetBstMidEndSepPunct{\mcitedefaultmidpunct}
{\mcitedefaultendpunct}{\mcitedefaultseppunct}\relax
\EndOfBibitem
\bibitem[Simha and Ramaswamy(2002)]{Simha2002}
R.~A. Simha and S.~Ramaswamy, \emph{Phys. Rev. Lett.}, 2002, \textbf{89},
  058101\relax
\mciteBstWouldAddEndPuncttrue
\mciteSetBstMidEndSepPunct{\mcitedefaultmidpunct}
{\mcitedefaultendpunct}{\mcitedefaultseppunct}\relax
\EndOfBibitem
\bibitem[Marchetti \emph{et~al.}(2013)Marchetti, Joanny, Ramaswamy, Liverpool,
  Prost, Rao, and Simha]{Marchetti2013}
M.~C. Marchetti, J.~F. Joanny, S.~Ramaswamy, T.~B. Liverpool, J.~Prost, M.~Rao
  and R.~A. Simha, \emph{Rev. Mod. Phys.}, 2013, \textbf{85}, 1143--1189\relax
\mciteBstWouldAddEndPuncttrue
\mciteSetBstMidEndSepPunct{\mcitedefaultmidpunct}
{\mcitedefaultendpunct}{\mcitedefaultseppunct}\relax
\EndOfBibitem
\bibitem[Prost \emph{et~al.}(2015)Prost, J{\"{u}}licher, and Joanny]{Prost2015}
J.~Prost, F.~J{\"{u}}licher and J.-F. Joanny, \emph{Nature Physics}, 2015,
  \textbf{11}, 111--117\relax
\mciteBstWouldAddEndPuncttrue
\mciteSetBstMidEndSepPunct{\mcitedefaultmidpunct}
{\mcitedefaultendpunct}{\mcitedefaultseppunct}\relax
\EndOfBibitem
\bibitem[Ramaswamy(2017)]{Ramaswamy2017}
S.~Ramaswamy, \emph{J. Stat. Mech. Theory Exp.}, 2017, \textbf{2017},
  054002\relax
\mciteBstWouldAddEndPuncttrue
\mciteSetBstMidEndSepPunct{\mcitedefaultmidpunct}
{\mcitedefaultendpunct}{\mcitedefaultseppunct}\relax
\EndOfBibitem
\bibitem[Doostmohammadi \emph{et~al.}(2018)Doostmohammadi, Ign{\'{e}}s-Mullol,
  Yeomans, and Sagu{\'{e}}s]{Doostmohammadi2018}
A.~Doostmohammadi, J.~Ign{\'{e}}s-Mullol, J.~M. Yeomans and F.~Sagu{\'{e}}s,
  \emph{Nat. Commun.}, 2018, \textbf{9}, 3246\relax
\mciteBstWouldAddEndPuncttrue
\mciteSetBstMidEndSepPunct{\mcitedefaultmidpunct}
{\mcitedefaultendpunct}{\mcitedefaultseppunct}\relax
\EndOfBibitem
\bibitem[Dafermos(1970)]{Dafermos1970}
C.~M. Dafermos, \emph{Q. J. Mech. Appl. Math}, 1970, \textbf{23}, 49--64\relax
\mciteBstWouldAddEndPuncttrue
\mciteSetBstMidEndSepPunct{\mcitedefaultmidpunct}
{\mcitedefaultendpunct}{\mcitedefaultseppunct}\relax
\EndOfBibitem
\bibitem[Imura and Okano(1973)]{Imura1973}
H.~Imura and K.~Okano, \emph{Phys. Lett. A}, 1973, \textbf{42}, 403--404\relax
\mciteBstWouldAddEndPuncttrue
\mciteSetBstMidEndSepPunct{\mcitedefaultmidpunct}
{\mcitedefaultendpunct}{\mcitedefaultseppunct}\relax
\EndOfBibitem
\bibitem[Pismen and Rodriguez(1990)]{Pismen1990}
L.~M. Pismen and J.~D. Rodriguez, \emph{Phys. Rev. A}, 1990, \textbf{42},
  2471--2474\relax
\mciteBstWouldAddEndPuncttrue
\mciteSetBstMidEndSepPunct{\mcitedefaultmidpunct}
{\mcitedefaultendpunct}{\mcitedefaultseppunct}\relax
\EndOfBibitem
\bibitem[Ryskin and Kremenetsky(1991)]{Ryskin1991}
G.~Ryskin and M.~Kremenetsky, \emph{Phys. Rev. Lett.}, 1991, \textbf{67},
  1574--1577\relax
\mciteBstWouldAddEndPuncttrue
\mciteSetBstMidEndSepPunct{\mcitedefaultmidpunct}
{\mcitedefaultendpunct}{\mcitedefaultseppunct}\relax
\EndOfBibitem
\bibitem[Denniston(1996)]{Denniston1996}
C.~Denniston, \emph{Phys. Rev. B}, 1996, \textbf{54}, 6272--6275\relax
\mciteBstWouldAddEndPuncttrue
\mciteSetBstMidEndSepPunct{\mcitedefaultmidpunct}
{\mcitedefaultendpunct}{\mcitedefaultseppunct}\relax
\EndOfBibitem
\bibitem[Radzihovsky(2015)]{Radzihovsky2015}
L.~Radzihovsky, \emph{Phys. Rev. Lett.}, 2015, \textbf{115}, 247801\relax
\mciteBstWouldAddEndPuncttrue
\mciteSetBstMidEndSepPunct{\mcitedefaultmidpunct}
{\mcitedefaultendpunct}{\mcitedefaultseppunct}\relax
\EndOfBibitem
\bibitem[Kats \emph{et~al.}(2002)Kats, Lebedev, and Malinin]{Kats2002}
E.~I. Kats, V.~V. Lebedev and S.~V. Malinin, \emph{J. Exp. Theor. Phys.}, 2002,
  \textbf{95}, 714--727\relax
\mciteBstWouldAddEndPuncttrue
\mciteSetBstMidEndSepPunct{\mcitedefaultmidpunct}
{\mcitedefaultendpunct}{\mcitedefaultseppunct}\relax
\EndOfBibitem
\bibitem[Sonnet(2005)]{Sonnet2005}
A.~M. Sonnet, \emph{Continuum Mech. Thermodyn.}, 2005, \textbf{17},
  287--295\relax
\mciteBstWouldAddEndPuncttrue
\mciteSetBstMidEndSepPunct{\mcitedefaultmidpunct}
{\mcitedefaultendpunct}{\mcitedefaultseppunct}\relax
\EndOfBibitem
\bibitem[Sonnet and Virga(2009)]{Sonnet2009}
A.~M. Sonnet and E.~G. Virga, \emph{Liquid Crystals}, 2009, \textbf{36},
  1185--1192\relax
\mciteBstWouldAddEndPuncttrue
\mciteSetBstMidEndSepPunct{\mcitedefaultmidpunct}
{\mcitedefaultendpunct}{\mcitedefaultseppunct}\relax
\EndOfBibitem
\bibitem[Giomi \emph{et~al.}(2013)Giomi, Bowick, Ma, and Marchetti]{Giomi2013}
L.~Giomi, M.~J. Bowick, X.~Ma and M.~C. Marchetti, \emph{Phys. Rev. Lett.},
  2013, \textbf{110}, 228101\relax
\mciteBstWouldAddEndPuncttrue
\mciteSetBstMidEndSepPunct{\mcitedefaultmidpunct}
{\mcitedefaultendpunct}{\mcitedefaultseppunct}\relax
\EndOfBibitem
\bibitem[Pismen(2013)]{Pismen2013}
L.~M. Pismen, \emph{Phys. Rev. E}, 2013, \textbf{88}, 050502\relax
\mciteBstWouldAddEndPuncttrue
\mciteSetBstMidEndSepPunct{\mcitedefaultmidpunct}
{\mcitedefaultendpunct}{\mcitedefaultseppunct}\relax
\EndOfBibitem
\bibitem[Giomi \emph{et~al.}(2014)Giomi, Bowick, Mishra, Sknepnek, and
  {Marchetti}]{Giomi2014}
L.~Giomi, M.~J. Bowick, P.~Mishra, R.~Sknepnek and M.~C. {Marchetti},
  \emph{Philos. Trans. Royal Soc. A}, 2014, \textbf{372}, 20130365\relax
\mciteBstWouldAddEndPuncttrue
\mciteSetBstMidEndSepPunct{\mcitedefaultmidpunct}
{\mcitedefaultendpunct}{\mcitedefaultseppunct}\relax
\EndOfBibitem
\bibitem[Zhang \emph{et~al.}(2018)Zhang, Kumar, Ross, Gardel, and
  de~Pablo]{Zhang2018}
R.~Zhang, N.~Kumar, J.~L. Ross, M.~L. Gardel and J.~J. de~Pablo, \emph{Proc.
  Natl. Acad. Sci. U.S.A.}, 2018, \textbf{115}, E124--E133\relax
\mciteBstWouldAddEndPuncttrue
\mciteSetBstMidEndSepPunct{\mcitedefaultmidpunct}
{\mcitedefaultendpunct}{\mcitedefaultseppunct}\relax
\EndOfBibitem
\bibitem[Cortese \emph{et~al.}(2018)Cortese, Eggers, and
  Liverpool]{Cortese2018}
D.~Cortese, J.~Eggers and T.~B. Liverpool, \emph{Phys. Rev. E}, 2018,
  \textbf{97}, 022704\relax
\mciteBstWouldAddEndPuncttrue
\mciteSetBstMidEndSepPunct{\mcitedefaultmidpunct}
{\mcitedefaultendpunct}{\mcitedefaultseppunct}\relax
\EndOfBibitem
\bibitem[Vromans and Giomi(2016)]{Vromans2016}
A.~J. Vromans and L.~Giomi, \emph{Soft Matter}, 2016, \textbf{12},
  6490--6495\relax
\mciteBstWouldAddEndPuncttrue
\mciteSetBstMidEndSepPunct{\mcitedefaultmidpunct}
{\mcitedefaultendpunct}{\mcitedefaultseppunct}\relax
\EndOfBibitem
\bibitem[Tang and Selinger(2017)]{Tang2017}
X.~Tang and J.~V. Selinger, \emph{Soft Matter}, 2017, \textbf{13},
  5481--5490\relax
\mciteBstWouldAddEndPuncttrue
\mciteSetBstMidEndSepPunct{\mcitedefaultmidpunct}
{\mcitedefaultendpunct}{\mcitedefaultseppunct}\relax
\EndOfBibitem
\bibitem[Shankar \emph{et~al.}(2018)Shankar, Ramaswamy, Marchetti, and
  Bowick]{Shankar2018}
S.~Shankar, S.~Ramaswamy, M.~C. Marchetti and M.~J. Bowick, \emph{Phys. Rev.
  Lett.}, 2018, \textbf{121}, 108002\relax
\mciteBstWouldAddEndPuncttrue
\mciteSetBstMidEndSepPunct{\mcitedefaultmidpunct}
{\mcitedefaultendpunct}{\mcitedefaultseppunct}\relax
\EndOfBibitem
\bibitem[Vertogen(1983)]{Vertogen1983}
G.~Vertogen, \emph{Z. Naturforsch. A}, 1983, \textbf{38}, 1273--1275\relax
\mciteBstWouldAddEndPuncttrue
\mciteSetBstMidEndSepPunct{\mcitedefaultmidpunct}
{\mcitedefaultendpunct}{\mcitedefaultseppunct}\relax
\EndOfBibitem
\bibitem[Vertogen and de~Jeu(1988)]{Vertogen1988}
G.~Vertogen and W.~H. de~Jeu, \emph{{Thermotropic Liquid Crystals,
  Fundamentals}}, Springer, 1988\relax
\mciteBstWouldAddEndPuncttrue
\mciteSetBstMidEndSepPunct{\mcitedefaultmidpunct}
{\mcitedefaultendpunct}{\mcitedefaultseppunct}\relax
\EndOfBibitem
\bibitem[Sonnet and Virga(2001)]{Sonnet2001}
A.~M. Sonnet and E.~G. Virga, \emph{Phys. Rev. E}, 2001, \textbf{64},
  031705\relax
\mciteBstWouldAddEndPuncttrue
\mciteSetBstMidEndSepPunct{\mcitedefaultmidpunct}
{\mcitedefaultendpunct}{\mcitedefaultseppunct}\relax
\EndOfBibitem
\bibitem[Sonnet and Virga(2012)]{Sonnet2012}
A.~M. Sonnet and E.~G. Virga, \emph{{Dissipative Ordered Fluids: Theories for
  Liquid Crystals}}, Springer, 2012\relax
\mciteBstWouldAddEndPuncttrue
\mciteSetBstMidEndSepPunct{\mcitedefaultmidpunct}
{\mcitedefaultendpunct}{\mcitedefaultseppunct}\relax
\EndOfBibitem
\bibitem[Doi(2011)]{Doi2011}
M.~Doi, \emph{J. Phys. Condens. Matter}, 2011, \textbf{23}, 284118\relax
\mciteBstWouldAddEndPuncttrue
\mciteSetBstMidEndSepPunct{\mcitedefaultmidpunct}
{\mcitedefaultendpunct}{\mcitedefaultseppunct}\relax
\EndOfBibitem
\bibitem[Stewart(2004)]{Stewart2004}
I.~W. Stewart, \emph{{The Static and Dynamic Continuum Theory of Liquid
  Crystals}}, Taylor \& Francis, 2004\relax
\mciteBstWouldAddEndPuncttrue
\mciteSetBstMidEndSepPunct{\mcitedefaultmidpunct}
{\mcitedefaultendpunct}{\mcitedefaultseppunct}\relax
\EndOfBibitem
\bibitem[Ryskin(1991)]{Ryskin1991b}
G.~Ryskin, \emph{J. Non-Newton. Fluid Mech.}, 1991, \textbf{39}, 207--210\relax
\mciteBstWouldAddEndPuncttrue
\mciteSetBstMidEndSepPunct{\mcitedefaultmidpunct}
{\mcitedefaultendpunct}{\mcitedefaultseppunct}\relax
\EndOfBibitem
\bibitem[Chakrabarty \emph{et~al.}(2013)Chakrabarty, Konya, Wang, Selinger,
  Sun, and Wei]{Chakrabarty2013}
A.~Chakrabarty, A.~Konya, F.~Wang, J.~V. Selinger, K.~Sun and Q.-H. Wei,
  \emph{Phys. Rev. Lett.}, 2013, \textbf{111}, 160603\relax
\mciteBstWouldAddEndPuncttrue
\mciteSetBstMidEndSepPunct{\mcitedefaultmidpunct}
{\mcitedefaultendpunct}{\mcitedefaultseppunct}\relax
\EndOfBibitem
\bibitem[Pieranski \emph{et~al.}(2016)Pieranski, Godinho, and
  {\v{C}}opar]{Pieranski2016a}
P.~Pieranski, M.~H. Godinho and S.~{\v{C}}opar, \emph{Phys. Rev. E}, 2016,
  \textbf{94}, 042706\relax
\mciteBstWouldAddEndPuncttrue
\mciteSetBstMidEndSepPunct{\mcitedefaultmidpunct}
{\mcitedefaultendpunct}{\mcitedefaultseppunct}\relax
\EndOfBibitem
\bibitem[Pieranski \emph{et~al.}(2016)Pieranski, {\v{C}}opar, Godinho, and
  Dazza]{Pieranski2016b}
P.~Pieranski, S.~{\v{C}}opar, M.~H. Godinho and M.~Dazza, \emph{Eur. Phys. J.
  E}, 2016, \textbf{39}, 121\relax
\mciteBstWouldAddEndPuncttrue
\mciteSetBstMidEndSepPunct{\mcitedefaultmidpunct}
{\mcitedefaultendpunct}{\mcitedefaultseppunct}\relax
\EndOfBibitem
\bibitem[Ravnik()]{Ravnik2018}
M.~Ravnik, personal communication\relax
\mciteBstWouldAddEndPuncttrue
\mciteSetBstMidEndSepPunct{\mcitedefaultmidpunct}
{\mcitedefaultendpunct}{\mcitedefaultseppunct}\relax
\EndOfBibitem
\bibitem[Norton \emph{et~al.}(2018)Norton, Baskaran, Opathalage, Langeslay,
  Fraden, Baskaran, and Hagan]{Norton2018}
M.~M. Norton, A.~Baskaran, A.~Opathalage, B.~Langeslay, S.~Fraden, A.~Baskaran
  and M.~F. Hagan, \emph{Physical Review E}, 2018, \textbf{97}, 012702\relax
\mciteBstWouldAddEndPuncttrue
\mciteSetBstMidEndSepPunct{\mcitedefaultmidpunct}
{\mcitedefaultendpunct}{\mcitedefaultseppunct}\relax
\EndOfBibitem
\bibitem[Ackerman \emph{et~al.}(2017)Ackerman, Boyle, and
  Smalyukh]{Ackerman2017}
P.~J. Ackerman, T.~Boyle and I.~I. Smalyukh, \emph{Nat. Commun.}, 2017,
  \textbf{8}, 673\relax
\mciteBstWouldAddEndPuncttrue
\mciteSetBstMidEndSepPunct{\mcitedefaultmidpunct}
{\mcitedefaultendpunct}{\mcitedefaultseppunct}\relax
\EndOfBibitem
\bibitem[Li \emph{et~al.}(2018)Li, Borshch, Xiao, Paladugu, Turiv,
  Shiyanovskii, and Lavrentovich]{Li2018}
B.-X. Li, V.~Borshch, R.-L. Xiao, S.~Paladugu, T.~Turiv, S.~V. Shiyanovskii and
  O.~D. Lavrentovich, \emph{Nat. Commun.}, 2018, \textbf{9}, 2912\relax
\mciteBstWouldAddEndPuncttrue
\mciteSetBstMidEndSepPunct{\mcitedefaultmidpunct}
{\mcitedefaultendpunct}{\mcitedefaultseppunct}\relax
\EndOfBibitem
\end{mcitethebibliography}
\bibliographystyle{rsc} 

\end{document}